\documentclass[twocol]{ametsoc}

\usepackage{dsfont} 
\usepackage{stmaryrd}

\journal{jas}

\bibpunct{(}{)}{;}{a}{}{,}

\title{Quantifying the annular mode dynamics in an idealized atmosphere}

    \authors{Pedram Hassanzadeh\correspondingauthor{Pedram Hassanzadeh}}
    \affiliation{Departments of Mechanical Engineering and Earth, Environmental and Planetary Sciences, Rice University, Houston, Texas}
    \email{pedram@rice.edu}

    \extraauthor{Zhiming Kuang}
    \extraaffil{\small{Department of Earth and Planetary Sciences and John A. Paulson School of Engineering and Applied Sciences, Harvard University, Cambridge, Massachusetts}}

\abstract{The linear response function (LRF) of an idealized GCM, the dry dynamical core with Held-Suarez physics, is used to accurately compute how eddy momentum and heat fluxes change in response to the zonal wind and temperature anomalies of the annular mode at the low-frequency limit. Using these results and  knowing the parameterizations of surface friction and thermal radiation in Held-Suarez physics, the contribution of each physical process (meridional and vertical eddy fluxes, surface friction, thermal radiation, and meridional advection) to the annular mode dynamics is quantified. Examining the quasi-geostrophic potential vorticity balance, it is shown that the eddy feedback is positive and increases the persistence of the annular mode by a factor of more than two. Furthermore, how eddy fluxes change in response to only the barotropic component of the annular mode, i.e., vertically averaged zonal wind (and no temperature) anomaly, is also calculated similarly. The response of eddy fluxes to the barotropic-only component of the annular mode is found to be drastically different from the response to the full (barotropic+baroclinic) annular mode anomaly. In the former, the barotropic governor effect significantly suppresses the eddy generation leading to a negative eddy feedback that decreases the persistence of the annular mode by nearly a factor of three. These results suggest that the baroclinic component of the annular mode anomaly, i.e., the increased low-level baroclinicity, is essential for the persistence of the annular mode, consistent with the baroclinic mechanism but not the barotropic mechanism proposed in the previous studies.} 

\begin{document}

\maketitle

%
\section{Introduction}\label{sec:intro}
In the extratropical circulation of both hemispheres, a long-recognized dominant pattern of variability at the intraseasonal to interannual timescales is the ``annular mode'', which is often derived from empirical orthogonal function (EOF) analysis of meteorological fields such as zonal-mean zonal wind ($\overline{u}$) or geopotential heights and has been known for decades \citep{kidson1988indices,thompson1998arctic,feldstein2000timescale}. The leading EOF (EOF1) of $\overline{u}$ features an equivalent-barotropic dipolar structure centered around the time-mean jet and describes north-south temporal fluctuations of the midlatitude westerlies \citep{nigam1990structure,hartmann1998wave,lorenz2001eddy,lorenz2003eddy}. The zonal index, the timeseries (principal component, PC) associated with the annular mode, is characterized by temporal persistence, i.e., characteristic timescale longer than the synoptic timescale: the year-around decorrelation ($e$-folding) timescale of the zonal index is $\sim 10$ days in both hemispheres \citep{thompson2014barotropic,thompson2015baroclinic} although there is a strong seasonal dependence with timescales of $\sim 15$ days in the Northern hemisphere (NH) in December-January and  $\sim 20$ days in the Southern hemisphere (SH) in November-December \citep[see Fig.~2 of][]{gerber2008annular}. 

It should be noted that recently, \citet{thompson2014barotropic}, \citet{thompson2014periodic}, and \citet{thompson2015baroclinic}  have found another dominant pattern of variability, called the ``baroclinic annular mode'' (BAM), in the extratropical circulation of both hemispheres. BAM emerges as the leading EOF of eddy kinetic energy and its PC exhibits quasi-periodicity at the timescales of $\sim 20-25$ ($\sim 25-30$) days in the NH (SH). \citet{boljka2018coupling} have offered some evidence for the coupling of the barotropic annular mode (EOF1 of $\overline{u}$) and the baroclinic annular mode. In this study, we focus on the former and simply refer to it as \emph{annular mode}.   

Annular modes also robustly emerge as the leading pattern of variability in a broad hierarchy of models, from stochastically forced barotropic models to comprehensive GCMs \citep[e.g.,][]{robinson1991dynamics,feldstein1998atmospheric,limpasuvan1999eddies,vallis2004mechanism,gerber2008testing,gerber2008annular,barnes2010effect,zurita2014impact, sheshadri2017propagating}. However, the annular modes simulated in GCMs are too persistent and the decorrelation timescales can be several times larger than the observed timescales \citep{gerber2008annular,gerber2008testing}. The problem of too-persistent annular modes, which exists across the hierarchy of idealized to comprehensive GCMs, is a matter of concern not only because a key aspect of a leading pattern of climate variability is not correctly simulated in GCMs, but also because the fluctuation-dissipation theorem (FDT) suggests that if the internal timescale of a system is overestimated, then the system's response to external forcings (e.g., changes in radiative forcing) is overestimated by the same factor as well \citep{gerber2008testing, ring08,shepherd2014}.  

To understand this component of climate variability and to potentially improve the GCMs, the persistence of the annular mode and its source have been extensively studied and debated in the past few decades. The persistence is often (but not always, see below) attributed to a positive eddy-jet feedback internal to the troposphere: eddies are altered by the zonal-mean anomaly (of the annular mode) such that the anomalous eddy fluxes reinforce the zonal-mean anomaly, thus increasing the persistence of the annular mode \citep[e.g.,][]{robinson1991dynamics,yu1993zonal,branstator1995organization,limpasuvan2000wave,robinson2000baroclinic,lorenz2001eddy,lorenz2003eddy}. \citet{lorenz2001eddy} formulated a simple linear feedback model for the zonal index and developed a statistical method to quantify the feedback. Using the reanalysis data, they found evidence for the existence of a positive eddy-jet feedback and quantified its magnitude in the SH and NH extratropical circulations \citep{lorenz2001eddy,lorenz2003eddy}. \citet{simpson2013southernII} have recently proposed another statistical framework to quantify the feedback in reanalysis data.  

The mechanism of this internal (to troposphere) eddy-jet feedback, however, has remained unclear. Some studies have argued for a \emph{barotropic} mechanism for the feedback: the annular mode's anomalous $\overline{u}$ changes the upper-tropospheric meridional shear, thus altering the upper-level wave propagation and leading to a positive eddy-jet feedback \citep[e.g.,][]{gerber2007eddy,chen2008tropospheric,nie2014quantifying,lorenz2014understanding}. Some other studies have supported a \emph{baroclinic} mechanism for the feedback: the annular mode's zonal-mean anomaly changes the lower-tropospheric baroclinicity, thus altering the low-level eddy generation and leading to a positive eddy-jet feedback \citep[e.g.,][]{robinson1996does,robinson2000baroclinic,lorenz2001eddy,robert2017positive,boljka2018coupling}. We emphasize that the two mechanisms are not mutually exclusive. We also emphasize again that the discussion here is concerned with the barotropic annular mode and the terms barotropic and baroclinic mechanisms should not be confused with barotropic/baroclinic annular modes. 

Even the source of the annular mode persistence is not a settled issue and a few studies have questioned the importance of the internal eddy-jet feedback \citep{feldstein1998atmospheric,byrne2016annular,byrne2017nonstationarity,byrne2018seasonal}. In fact, \citet{byrne2016annular} have shown that the statistical methods, such as the ones developed in \citet{lorenz2001eddy} and \citet{simpson2013southernII}, cannot distinguish between an internal eddy-jet feedback and a low-frequency external forcing, casting doubt on the evidence for the internal feedback in the SH reanalysis data. They have further argued that this external forcing is a result of the stratospheric variability \citep{byrne2017nonstationarity,byrne2018seasonal}.         
                 
The major challenge in understanding and quantifying the role of eddies in the annular mode dynamics is the lack of a complete theory that allows us to compute changes in eddy fluxes in terms of changes in the mean-flow. Specifically, we need to calculate how eddy momentum and heat fluxes change in response to the anomalous $(\overline{u},\overline{T})$ of the annular mode in the low-frequency ($\sim$ steady-state) limit, see, e.g., the discussion in \citet{robinson2000baroclinic} and in particular the last sentence\footnote{``In the absence of a complete theory that allows
us to calculate eddy quantities in terms of the mean flow, the necessary next step is a careful analysis of the observed relationships between eddy fluxes and zonally averaged fields.''}. Recently, \citet{hassanzadeh2016linear} have introduced a framework, based on using the linear response functions (LRFs), that can help with addressing these types of eddy-mean flow interaction problems in GCMs. 

Applying this framework to an idealized GCM, the dry dynamical core with Held-Suarez physics \citep{held1994proposal}, \citet{ma2017quantifying} showed unequivocally that there is an internal positive eddy-jet feedback in the annular mode dynamics of this idealized atmosphere, accurately quantified the feedback magnitude, and developed a new statistical method, based on low-pass filtering, to compute the feedback from data. The estimates of feedback magnitude in the GCM and SH reanalysis data were shown to be more robust with respect to the free parameters in the low-pass filtering method of \citet{ma2017quantifying} compared to in other statistical methods. However, the problem of isolating internal feedback from external forcing in reanalysis data using statistical methods still exists.

The work of \citet{ma2017quantifying} was limited to the zonal index, a scalar quantity based on zonal wind. Here, we shall extend the analysis to the full latitude-pressure plane and include both zonal wind and temperature to further explore the detailed physical processes that produce the positive feedback. More specifically, in the current study we aim to use the LRF framework of \citet{hassanzadeh2016linear} and the same idealized setup as \citet{ma2017quantifying} to
\begin{enumerate}
\item Quantify the contribution of different physical processes (eddy momentum flux, eddy heat flux, surface friction, thermal radiation, and meridional advection) to the annular mode dynamics,
\item Quantify the contribution of the eddy-jet feedback to the annular mode persistence,
\item Examine the role of baroclinic and barotropic mechanisms in the feedback dynamics.
\end{enumerate}

The remainder of the paper is structured as follows. The idealized GCM and its annular mode are discussed in section~\ref{sec:gcm}. The LRF framework and simulations for quantifying the eddy-jet feedback are presented in section \ref{sec:forced}. Quantifying different physical processes and the annular mode budget are discussed in sections~\ref{sec:quant} and \ref{sec:res}, respectively. Discussions and summery are in section~\ref{sec:sum}.

\section{The Control Simulation and its Annular Mode}\label{sec:gcm}
We use the Geophysical Fluid Dynamics Laboratory (GFDL) dry dynamical core, which is a pseudo-spectral GCM that solves the primitive equations on sigma ($\sigma$) levels. We use the Held-Suarez physics with the configuration and physical parameters identical to the ones described in detail in \citet{held1994proposal}. Briefly, the model is forced by Newtonian relaxation of temperature to a prescribed equinoctial radiative-equilibrium state $T_{eq}(\phi,p)$ (Eq.~(\ref{eq:Teq})) with specified relaxation rate $k_T(\phi,p)$ (Eq.~(\ref{eq:kT})). $\phi$ and $p$ are latitude and pressure, respectively. Rayleigh drag with a prescribed damping rate $k_v(\phi,p)$ (Eq.~(\ref{eq:kv})) is used to remove momentum from the low levels ($\sigma<0.7$); $\nabla^8$ hyper-diffusion is used to remove enstrophy at small scales. A T63 spectral resolution with 40 equally spaced sigma levels and $15$-min time steps are used to solve the equations. The above setup is referred to as control (CTL) hereafter. The CTL simulation consists of ten ensemble runs, each 45000 days with 6-hourly outputs. The first 500~days of each run are discarded for spin-up and given the hemispherical symmetry of the model, the data of the northern and southern hemispheres are used together in the calculations of the time averages and EOF analysis.

The dry dynamical core GCM has been widely used to study the extratropical circulation and its low-frequency variability \citep{gerber2008testing,chen2009quantifying,pedram14,sheshadri2018vertical,ronalds2018barotropic} and is deemed a key member of the hierarchy of \emph{elegant} climate models \citep{held2005gap,jeevanjee2017perspective}. The setup of the idealized GCM used here does not have a dynamically active stratosphere and provides a minimal primitive-equation model to study the dynamics of the annular mode internal to the troposphere.

\begin{figure*}[t]
\centerline{\includegraphics[width=1\textwidth]{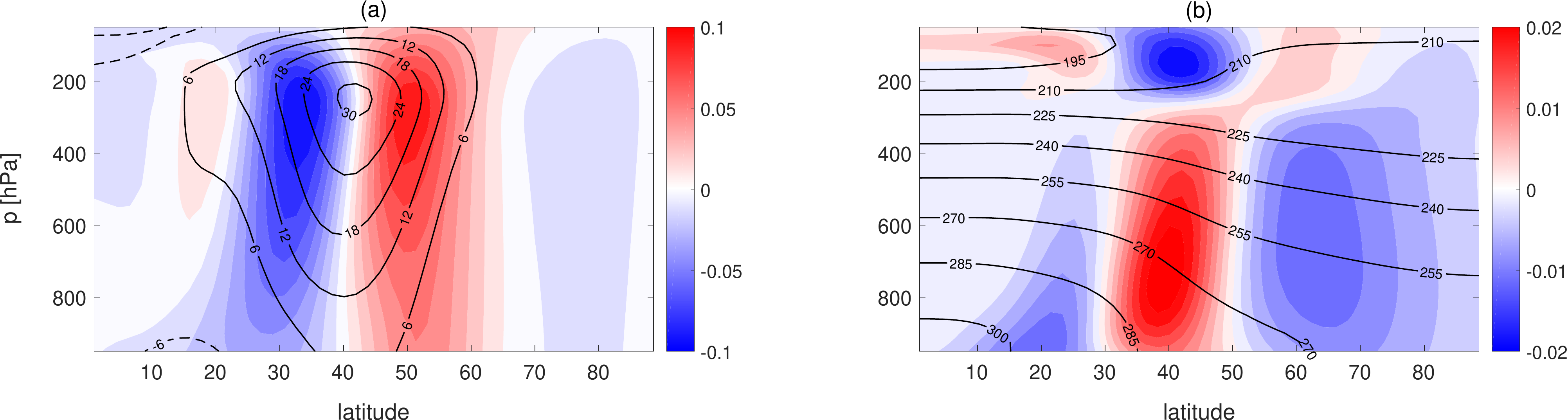}}
\caption{The climatology (contour) and annular mode (i.e., EOF1; shading) of the CTL: (a) zonal wind [m~s$^{-1}$] and (b) temperature [K]. The interval of contour lines is $6$~m s$^{-1}$ for zonal wind and $15$~K for temperature. Dashed lines are for negative values.}
\label{fig:1}
\end{figure*}

The model's annular mode is obtained from the EOF analysis of daily averaged $\overline{u}$ which provides $\overline{u}_\mathrm{EOF1}$ and zonal index (PC1). $\overline{T}_\mathrm{EOF1}$ is obtained from regressing $\overline{T}$ on the zonal index. Figure~\ref{fig:1} shows $\overline{u}_\mathrm{EOF1}$ and $\overline{T}_\mathrm{EOF1}$ of this model, which closely resemble the annular mode in reanalysis data of the NH and SH; see Figs.~2 in \citet{thompson2015baroclinic} and \citet{thompson2014barotropic}. \citet{ma2017quantifying} compared the spectra, auto-correlation, and cross-correlation of the anomalous zonal-mean zonal wind and eddy momentum flux  projected onto EOF1 in this model and the SH reanalysis data, and found that the spatio-temporal properties of this GCM's annular mode overall agree reasonably well with the observed ones, with the exception of a too-persistent decorrelation timescale of the zonal index. As discussed earlier, this is a common problem among GCMs. 

\section{The Forced Simulations and Eddy Feedbacks}\label{sec:forced}
\paragraph{EXP1: Barotropic+Baroclinic Components of Annular Mode --}\label{sec:forced_BCT}
To accurately quantify the response of eddy momentum and heat fluxes to the annular mode's zonal wind and temperature anomaly $(\overline{u}_\mathrm{EOF1},\overline{T}_\mathrm{EOF1})$, we have used the framework that is introduced in \citet{hassanzadeh2016linear} and employed in \citet{ma2017quantifying}. In this framework, a second simulation (referred to as EXP1 hereafter) is conduced in which a time-invariant, zonally symmetric forcing $\overline{\boldsymbol{f}}_\mathrm{EXP1}$ of $\overline{u}$ and $\overline{T}$ is added to the GCM such that the time-mean responses of $\overline{u}$ and $\overline{T}$ compared to CTL almost match the $\overline{u}_\mathrm{EOF1}$ and $\overline{T}_\mathrm{EOF1}$ of CTL. The forcing $\overline{\boldsymbol{f}}_\mathrm{EXP1}$ is calculated using the GCM's linear response function $\pmb{\mathsf{L}}$ as $\overline{\boldsymbol{f}}_\mathrm{EXP1} = - \pmb{\mathsf{L}} \, \overline{\boldsymbol{x}}_\mathrm{EOF1}$ where $\overline{\boldsymbol{x}}_\mathrm{EOF1}$ is a vector containing $\overline{u}_\mathrm{EOF1}$ and $\overline{T}_\mathrm{EOF1}$ (this vector is scaled such that the maximum of $\overline{u}_\mathrm{EOF1}$ is $0.1$~m s$^{-1}$). We use $\pmb{\mathsf{L}}$ of the exact same setup of this GCM that was previously calculated by \citet{hassanzadeh2016linear} using the Green's function method. EXP1 is essentially the same as test~3 in \citet{hassanzadeh2016linear} and the forced experiment in \citet{pedram15}. 

\begin{figure*}[t]
\centerline{\includegraphics[width=1\textwidth]{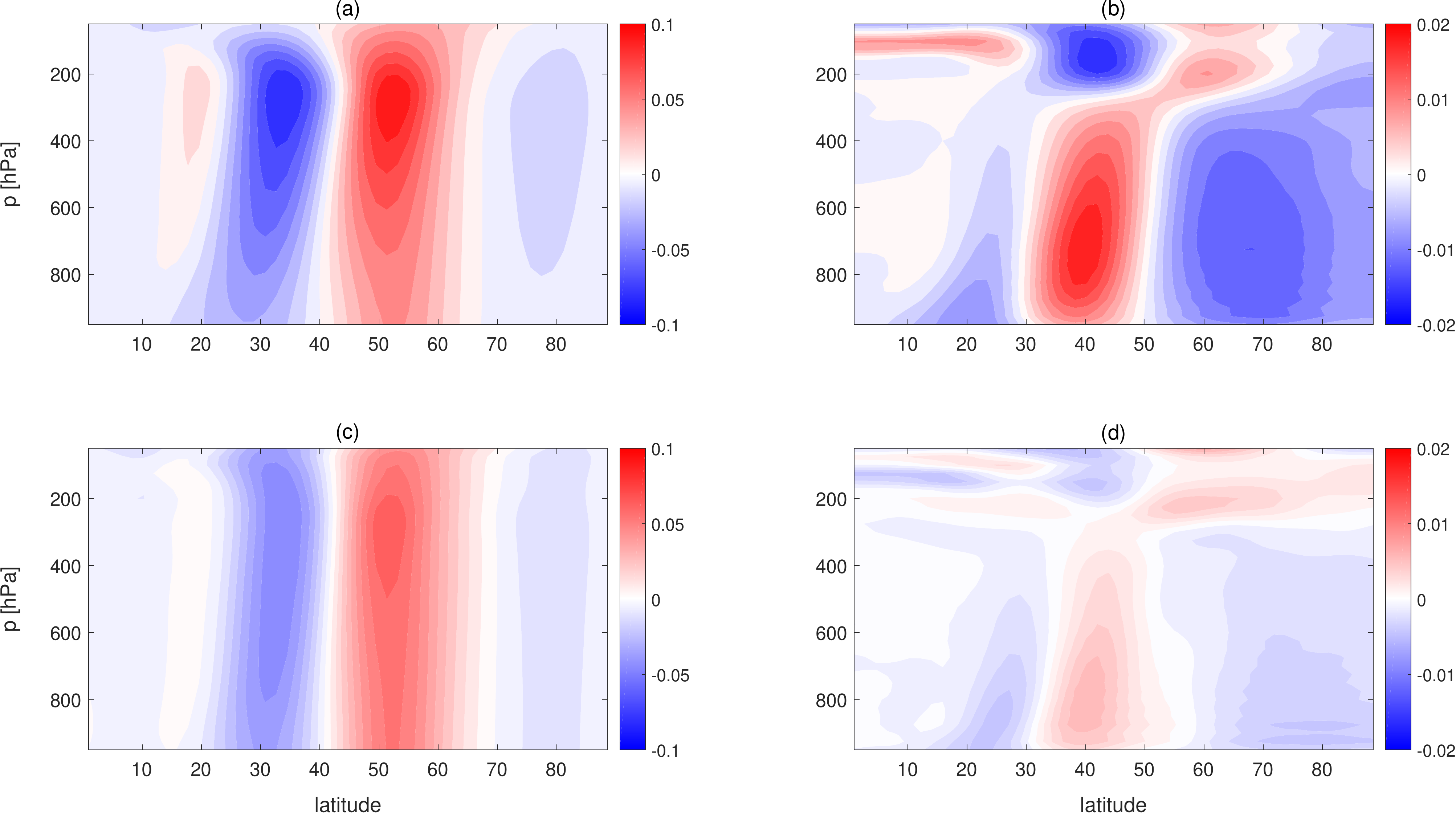}}
\caption{The forced simulations. Full (barotropic+baroclinic) annular mode anomaly (EXP1): (a) $\langle \overline{u}  \rangle_\mathrm{EXP1} -\langle \overline{u}  \rangle_\mathrm{CTL}$~[m s$^{-1}$] and (b) $\langle \overline{T}  \rangle_\mathrm{EXP1} -\langle \overline{T}  \rangle_\mathrm{CTL}$~[K]. Only the barotropic component of the annular mode anomaly (EXP2) (c) $\langle \overline{u}  \rangle_\mathrm{EXP2} -\langle \overline{u}  \rangle_\mathrm{CTL}$~[m s$^{-1}$] and (d) $\langle \overline{T}  \rangle_\mathrm{EXP2} -\langle \overline{T}  \rangle_\mathrm{CTL}$~[K].} 
\label{fig:2}
\end{figure*}

Similar to CTL, EXP1 consists of ten ensemble runs, each 45000 days. With $\langle \; \rangle$ indicating time and ensemble averaging, Figures~\ref{fig:2}(a)-(b) show $\langle \overline{u}  \rangle_\mathrm{EXP1} -\langle \overline{u}  \rangle_\mathrm{CTL}$ and $\langle \overline{T}  \rangle_\mathrm{EXP1} -\langle \overline{T}  \rangle_\mathrm{CTL}$, which agree well with the EOF1 of CTL (Figures~\ref{fig:1}(a)-(b)). The long datasets of CTL and EXP1 (around one million days each) are needed to ensure that the weak time-mean response emerges from the noise. The distinction between the patterns in Figures~\ref{fig:1}(a)-(b) and \ref{fig:2}(a)-(b) should be highlighted: the former shows the pattern of variability that explains the most variance in CTL while the latter shows the difference between the climatologies of EXP1 and CTL. 

\begin{figure*}[t]
\centerline{\includegraphics[width=1\textwidth]{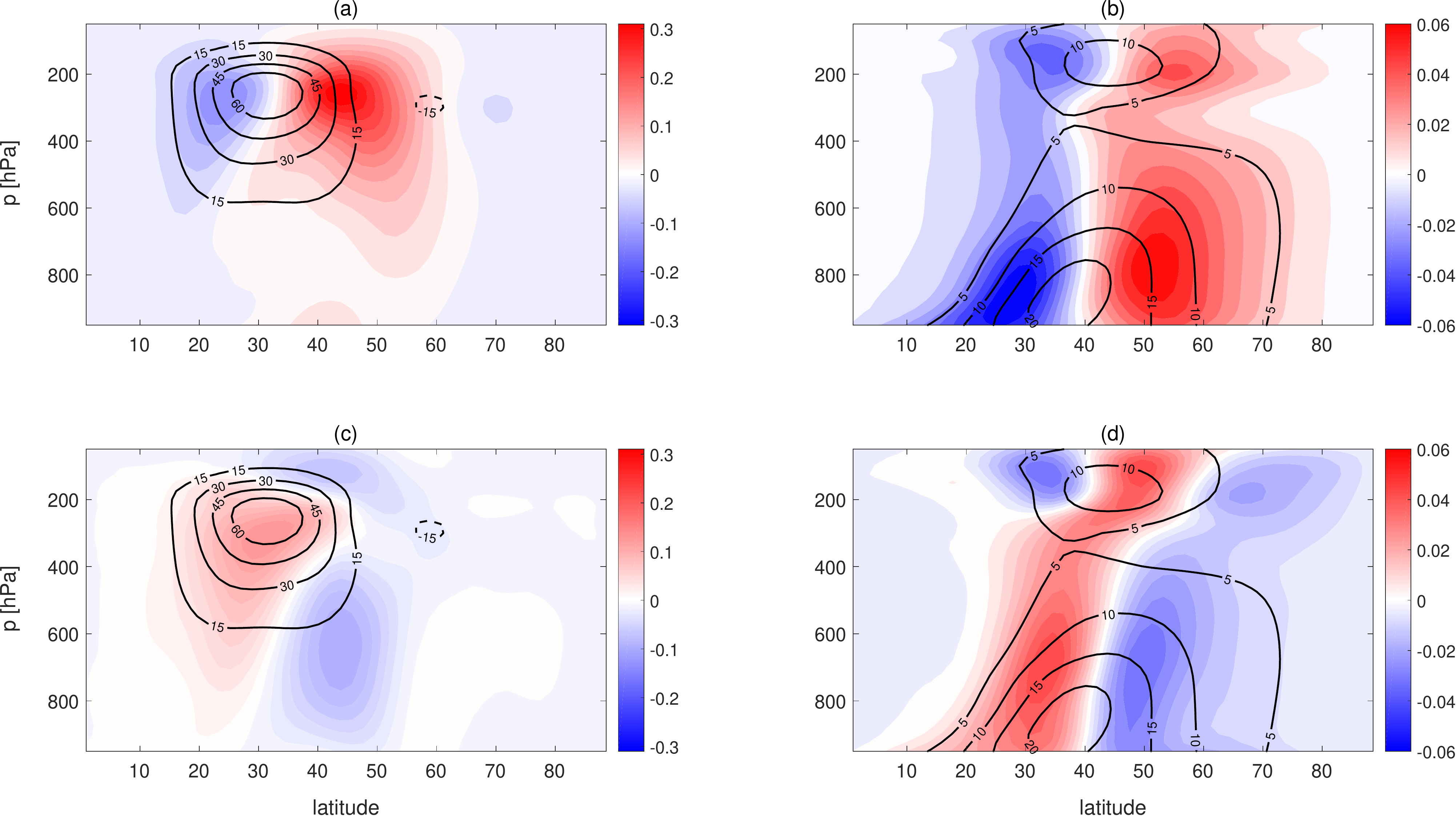}}
\caption{Eddy momentum and heat fluxes: climatology of CTL (contour) and response to the annular mode pattern (shading). Response to the full (barotropic+baroclinic) annular mode anomaly (EXP1): (a) $\langle \overline{u'v'}  \rangle_\mathrm{EXP1} -\langle \overline{u'v'}  \rangle_\mathrm{CTL}$~m$^2$ s$^{-2}$ and (b) $\langle \overline{v'T'}  \rangle_\mathrm{EXP1} -\langle \overline{v'T'}  \rangle_\mathrm{CTL}$~m s$^{-1}$ K. Response to only the barotropic component of the annular mode anomaly (EXP2): (c) $\langle \overline{u'v'}  \rangle_\mathrm{EXP2} -\langle \overline{u'v'}  \rangle_\mathrm{CTL}$ [m$^2$ s$^{-2}$] and (d) $\langle \overline{v'T'}  \rangle_\mathrm{EXP2} -\langle \overline{v'T'}  \rangle_\mathrm{CTL}$~[m s$^{-1}$ K]. The interval of contour lines is $15$~m$^2$ s$^{-2}$ for eddy momentum flux and $5$~m s$^{-1}$~K for eddy heat flux. Dashed lines are for negative values. } 
\label{fig:3}
\end{figure*}

Figures~\ref{fig:3}(a)-(b) show the differences between climatology of eddy momentum and heat fluxes in EXP1 and CTL: $\langle \overline{u'v'}  \rangle_\mathrm{EXP1} -\langle \overline{u'v'}  \rangle_\mathrm{CTL}$ and $\langle \overline{v'T'}  \rangle_\mathrm{EXP1} -\langle \overline{v'T'}  \rangle_\mathrm{CTL}$, where primes indicate deviation from zonal mean. Because the forcing added in EXP1, $\overline{\boldsymbol{f}}_\mathrm{EXP1}$, is time-invariant and zonally symmetric, changes in the climatology of eddy fluxes between EXP1 and CTL are solely due to changes in the climatology of $\overline{u}$ and $\overline{T}$ between EXP1 and CTL; see \citet{hassanzadeh2016linear} and \citet{ma2017quantifying} for further discussions. Therefore, Figures~\ref{fig:3}(a)-(b) show the anomalous eddy fluxes in response to the annular mode anomaly (Figures~\ref{fig:2}(a)-(b)) at the steady-state limit, finding which was the main difficulty in quantifying the role of eddies in annular mode dynamics as discussed in section~\ref{sec:intro}. In fact, following the linear feedback framework of \citet{lorenz2001eddy}, \citet{ma2017quantifying} calculated the eddy momentum forcing $F_{eddy}=-\partial \overline{u'v'}/\partial y$ from Figure~\ref{fig:3}(a) and used $\overline{u}_\mathrm{AM}=\langle \overline{u}  \rangle_\mathrm{EXP1} -\langle \overline{u}  \rangle_\mathrm{CTL}$ from Figure~\ref{fig:2}(a) to compute the feedback strength 
\begin{eqnarray}
b=\llbracket F_{eddy}  \overline{u}_\mathrm{AM} \rrbracket/\llbracket\overline{u}_\mathrm{AM}  \overline{u}_\mathrm{AM} \rrbracket=+0.14 \; \mathrm{day}^{-1},
\label{eq:feed}
\end{eqnarray}
where $\llbracket \; \rrbracket$ indicates spatially weighted domain averaging. Note that all calculations in this paper are conducted in the spherical coordinates, but $\partial /\partial y$ is used for convenience. The statistical methods of \citet{lorenz2001eddy} and \citet{simpson2013southernII} yield a similar but slightly lower estimate for $b$ ($0.11-0.13$~day$^{-1}$ depending on the chosen free parameter) and the low-pass filtering method robustly gives the same estimate of $+0.14$~day$^{-1}$ for low-pass timescales larger than 200 days \citep[see Fig. 8 in][]{ma2017quantifying}.        

We highlight that the eddy fluxes in this paper are calculated from $6$-hourly outputs to capture the medium-scale waves, which have timescales shorter than two days \citep{sato2000global}. Despite their weak climatological amplitudes, the medium-scale waves are strongly modified by the annular mode anomaly and their fluxes have a substantial contribution to the annular mode dynamics \citep{kuroda2011role}, e.g., the eddy momentum forcing on the annular mode in the low-frequency limit can be underestimated by around $50\%$ if the eddy fluxes are calculated from daily-averaged wind \citep[][Fig. 10]{ma2017quantifying}. 

In addition to the meridional eddy fluxes, we have also calculated the responses of vertical fluxes: $\langle \overline{u'\omega'}\rangle_\mathrm{EXP1}-\langle\overline{u'\omega'} \rangle_\mathrm{CTL}$ and $\langle \overline{\omega'T'}\rangle_\mathrm{EXP1}-\langle\overline{\omega'T'} \rangle_\mathrm{CTL} $ where $\omega$ is the pressure velocity. 

In summary, the above analysis of CTL and EXP1 confirms the existence of a positive eddy-jet feedback and quantifies its magnitude in this idealized atmosphere (as originally reported in \citet{ma2017quantifying}). This analysis also allows us to compute the contributions of the meridional and vertical eddy momentum and heat fluxes to the annular mode dynamics in sections~\ref{sec:res} in order to address objectives~1 and 2 that were mentioned in section~\ref{sec:intro}. To better address objective 3, we conduct a second experiment EXP2 described below. We emphasize that in EXP2, the aim is to quantify the \emph{net} effect of change in meridional shear, which not only affects upper-level wave propagation, but also low-level wave generation; this distinction is further discussed at the end of this section. 


\paragraph{EXP2: Only Barotropic Component of Annular Mode --}\label{sec:forced_BT}
To further investigate the effect of the annular mode anomaly on the eddy heat and momentum fluxes, a second forced simulation (EXP2) is conduced in which the added forcing $\overline{\boldsymbol{f}}_\mathrm{EXP2}$ is meant to produce a time-mean response in zonal wind that matches the {\it barotropic} component of $\overline{u}_\mathrm{EOF1}$. To find this forcing, first we have calculated the mass-weighted vertical average of $\overline{u}_\mathrm{EOF1}$ at each latitude, which along with zero temperature anomaly, constitutes $\overline{\boldsymbol{x}}_\mathrm{BTRO}=(\overline{u}_\mathrm{BTRO},0)$. We then calculated $\overline{\boldsymbol{f}}_\mathrm{EXP2} = - \pmb{\mathsf{L}} \, \overline{\boldsymbol{x}}_\mathrm{BTRO}$. As before, EXP2 consists of ten ensemble runs, each 45000 days. Figures~\ref{fig:2}(c)-(d) show $\langle \overline{u}  \rangle_\mathrm{EXP2} -\langle \overline{u}  \rangle_\mathrm{CTL}$ and $\langle \overline{T}  \rangle_\mathrm{EXP2} -\langle \overline{T}  \rangle_\mathrm{CTL}$. The barotropic component of $\overline{u}_\mathrm{EOF1}$ is reproduced reasonably well (compare with panel~(a)), although there are some small variations with pressure, due to the inaccuracies in $\pmb{\mathsf{L}}$. As a result, there is also a small temperature anomaly (panel (d)). Overall, EXP2 captures the barotropic component of EOF1 adequately for the purpose of our analysis.

Figures~\ref{fig:3}(c)-(d) show the differences between the climatology of eddy momentum and heat fluxes in EXP2 and CTL. As discussed earlier, these differences are the response of eddy fluxes to the (mostly) barotropic component of the annular mode shown in Figures~\ref{fig:2}(c)-(d). There is a striking difference between the response of the eddy fluxes to the (mostly) \emph{barotropic-only component} of the annular mode anomaly and to the \emph{full (barotropic+baroclinic)} annular mode anomaly as can be seen by comparing the two rows of Figure~\ref{fig:3}. 

As pointed out in \citet{ma2017quantifying}, the change in eddy fluxes in response to the full annular mode anomaly is consistent with the baroclinic mechanism discussed in \citet{robinson2000baroclinic} and \citet{lorenz2001eddy}: increased (reduced) baroclinicity around the latitude of maximum (minimum ) $\overline{u}_\mathrm{EOF1}$ anomaly, $50^\mathrm{o}$ ($30^\mathrm{o}$) in Figure~\ref{fig:2}(a), which is due to increased (reduced) meridional temperature gradient around the same latitude, see Figure~\ref{fig:2}(b), leads to increased (reduced) eddy generation and $\overline{v'T'}$ at low levels with a maximum (minimum) around $50^\mathrm{o}$ ($30^\mathrm{o}$), see Figure~\ref{fig:3}(b). The upward (downward) wave propagation around $50^\mathrm{o}$ ($30^\mathrm{o}$) combined with the prevalence of equatorward wave propagation at upper levels due to spherical geometry \citep{balasubramanian1997role} lead to a maximum of $\overline{u'v'}$ between $40^\mathrm{o}-50^\mathrm{o}$ at upper levels, see Figure~\ref{fig:3}(a). The resulting eddy forcing $-\partial \overline{u'v'}/\partial y$ has its maximum (minimum) at upper levels around the latitude of maximum (minimum) $\overline{u}_\mathrm{EOF1}$ anomaly $50^\mathrm{o}$ ($30^\mathrm{o}$). This pattern of eddy forcing leads to reinforcement of the annular mode anomaly with the feedback magnitude calculated in Eq.~(\ref{eq:feed}). 

The change in eddy fluxes in response to only the barotropic component of the annular mode anomaly is quiet different and in particular, the response of $\overline{v'T'}$ is opposite: around the latitude of maximum (minimum) $\overline{u}_\mathrm{BTRO}$ anomaly, which is the same as the corresponding latitude for the $\overline{u}_\mathrm{EOF1}$ anomaly, there is reduced (increased) eddy generation and $\overline{v'T'}$ at low levels with a minimum (maximum) around $50^\mathrm{o}$ ($30^\mathrm{o}$), see Figure~\ref{fig:3}(d). This is accompanied by reduced (increased) $\overline{u'v'}$ below 400~hPa around $45^\mathrm{o}$ (above 500~hPa around $30^\mathrm{o}$), see Figure~\ref{fig:3}(c). The resulting eddy forcing $-\partial \overline{u'v'}/\partial y$ has its minimum (maximum) throughout the troposphere around the latitude of maximum (minimum) $\overline{u}_\mathrm{BTRO}$ anomaly. This pattern of eddy forcing leads to weakening of the annular mode anomaly with the negative feedback magnitude calculated following Eq.~(\ref{eq:feed}) to be $\sim -0.1$~day$^{-1}$. 

The substantial change in  $\overline{v'T'}$ (Figure~\ref{fig:3}(d)) is in spite of almost no change in baroclinicity (Figure~\ref{fig:2}(d)); in fact, the little change in $\overline{T}$ still resembles the one in Figure~\ref{fig:2}(b) and is expected to lead to a response that has the same pattern, but much weaker amplitude, compared to the one shown in Figure~\ref{fig:3}(b). However, that is not the case. Instead, the response of $\overline{v'T'}$ (and the resulting change in $\overline{u'v'}$) is consistent with the barotropic governor effect, i.e., stronger meridional shear can restrict the structure of the baroclinic waves and reduce the conversion of potential energy to eddy kinetic energy thus suppressing the eddy generation in baroclinic instability \citep{james1986concerning,james1987suppression,nakamura1993momentum}. The meridional shear associated with the $\overline{u}_\mathrm{BTRO}$ anomaly decreases (increases) the total meridional shear around $30^\mathrm{o}$ ($50^\mathrm{o}$) leading to the intensification (suppression) of eddy generation and $\overline{v'T'}$ around these latitudes.                 

As mentioned earlier, some studies have suggested that changes in the upper-level meridional shear lead to an eddy forcing that reinforces the annular mode anomaly. The analysis presented above is not refuting that mechanism, but showing that the net effect of the change in meridional shear on upper-level wave propagation and low-level wave generation is a negative feedback, likely because the barotropic governor mechanism prevails over any potential increase in upper-level wave propagation. Separating these effects is beyond the scope of this paper and should be examined in future studies. However, to better address objective 3, we examine the budget for both full and barotropic-only  annular mode anomalies in section~\ref{sec:res}. 

\section{Quantifying Physical Processes} \label{sec:quant}
Knowing the meridional and vertical eddy momentum and heat fluxes (section~\ref{sec:forced}) and the GCM's Rayleigh drag and the Newtonian cooling parameterizations (Appendix~A), the contribution of each physical process to the annular mode dynamics (i.e., the tendency balance of $\overline{u}_\mathrm{AM}$ and $\overline{T}_\mathrm{AM}$) can be quantified. However, in addition to the direct contributions, these processes can each also drive a secondary (meridional) circulation and indirectly affect the tendency balance of $\overline{u}_\mathrm{AM}$ and $\overline{T}_\mathrm{AM}$. For example, the surface friction forcing $- k_v \overline{u}_\mathrm{AM}$ affects the low-level baroclinicity through adiabatic warming resulting from the induced secondary circulation (see below). The combination of direct and indirect contributions from a forcing has been referred to as the ``Eliassen response'' \citep[][section~5]{ring08} and can be calculated by solving the balanced ``Eliassen problem'' \citep{eliassen1951slow}.    

To further illustrate this concept and the role of each physical process, we look at the zonally averaged primitive equations for annular mode anomalies of zonal wind $\overline{u}_\mathrm{AM}$ and potential temperature $\overline{\theta}_\mathrm{AM}$. Defining the CTL time-mean (background) flow $(\overline{U},\overline{V},\overline{\Omega},\overline{\Theta})=(\langle \overline{u} \rangle,\langle \overline{v} \rangle, \langle \overline{\omega} \rangle,\langle \overline{\theta} \rangle)_\mathrm{CTL}$, radius of earth $a$, and Coriolis parameter $f$, these equations become \citep{simpson2013southernII,hassanzadeh2016linear}:
\begin{eqnarray}
\frac{\partial \overline{u}_\mathrm{AM}}{\partial t} &=& -\overline{V} \frac{1}{a \cos{\phi}}\frac{\partial (\overline{u}_\mathrm{AM}\cos{\phi})}{\partial \phi }-\overline{\Omega}\frac{\partial \overline{u}_\mathrm{AM}}{\partial p}   \nonumber \\
&&+ \overline{v}_\mathrm{AM}\left[f-\frac{1}{a \cos{\phi}}\frac{\partial ((\overline{U}+ \overline{u}_\mathrm{AM})\cos{\phi})}{\partial \phi }\right]
\nonumber \\
&&-\overline{\omega}_\mathrm{AM}\frac{\partial (\overline{U}+ \overline{u}_\mathrm{AM})}{\partial p}  \nonumber \\
&&-\frac{1}{a \cos^2{\phi}}\frac{\partial (\overline{u'v'}_\mathrm{AM}\cos^2{\phi})}{\partial \phi}-\frac{\partial \overline{u'\omega'}_\mathrm{AM}}{\partial p}  \nonumber \\
&&- k_v \overline{u}_\mathrm{AM}, \label{eq:u} 
\end{eqnarray}
\begin{eqnarray}
\frac{\partial \overline{\theta}_\mathrm{AM}}{\partial t} &=& -\overline{V} \frac{1}{a}\frac{\partial \overline{\theta}_\mathrm{AM}}{\partial \phi }-\overline{\Omega}\frac{\partial \overline{\theta}_\mathrm{AM}}{\partial p}   \nonumber \\
&& -\overline{v}_\mathrm{AM} \frac{1}{a}\frac{\partial (\overline{\Theta}+\overline{\theta}_\mathrm{AM})}{\partial \phi } \nonumber \\
&& -\overline{\omega}_\mathrm{AM}\frac{\partial (\overline{\Theta}+\overline{\theta}_\mathrm{AM})}{\partial p}   \nonumber \\
&&-\frac{1}{a}\frac{\partial \overline{v'\theta'}_\mathrm{AM}}{\partial \phi}-\frac{\partial \overline{\omega'\theta'}_\mathrm{AM}}{\partial p}  \nonumber \\
&&- k_T \overline{\theta}_\mathrm{AM}. \label{eq:T}
\end{eqnarray}
On the right-hand side of Eq.~(\ref{eq:u}) (Eq.~(\ref{eq:T})), the two terms in the first line represent the advection of the annular mode zonal wind (temperature) anomaly by the meridional circulation of the background flow, the terms in the fourth line represent the anomalous meridional and vertical eddy momentum (heat) flux divergence, and the term in the last line represents the Rayleigh drag on the zonal wind anomaly (Newtonian cooling of the temperature anomaly). The terms in the second and third lines of both equations represent the total effect of the induced secondary circulations, which contribute to the tendency budget by meridionally advecting the total zonal wind and potential temperature. Because we aim to quantify the contribution of each physical process to the annular mode dynamics, we need to compute the secondary circulation and the resulting total tendency in Eqs.~(\ref{eq:u}) and (\ref{eq:T}) from each individual process. 

This can be achieved by solving the balanced Eliassen problem: the terms on the second and third lines of Eqs.~(\ref{eq:u}) and (\ref{eq:T}) are linearized around $(\overline{U},\overline{\Theta})$ (which is justified given that max$(\overline{u}_\mathrm{AM})=0.1$~m s$^{-1}$ and max$(\overline{U})\sim 30$~m s$^{-1}$) and the two equations are combined assuming gradient-wind balance (which eliminates the left-hand sides) and setting $(\overline{V},\overline{\Omega})=0$, resulting in a linear, prognostic equation for $(\overline{v},\overline{\omega})$ \citep[see Appendix A in][]{hassanzadeh2016linear}. The induced secondary circulation from each forcing in Eqs.~(\ref{eq:u}) and (\ref{eq:T}), e.g., $- k_v \overline{u}_\mathrm{AM}$, can now be computed by only including that term in the linear, prognostic equation and solving for $(\overline{v},\overline{\omega})$. The formulation and development of a solver for the Eliassen problem have been described in detail in appendix~A of \citet{ring08}. However, with little effort, the idealized GCM itself can be turned into an axisymmetric solver for this purpose, as described in Appendix~B. Using this solver, below we compute the induced circulation and the total contribution of each process to the tendency of the full annular mode anomaly. Similar analysis for the barotropic-only component of the annular mode anomaly is presented in Appendix~C.                           

\subsection{Surface Friction} \label{subsec:fric}
\begin{figure*}[t]
\centerline{\includegraphics[width=1\textwidth]{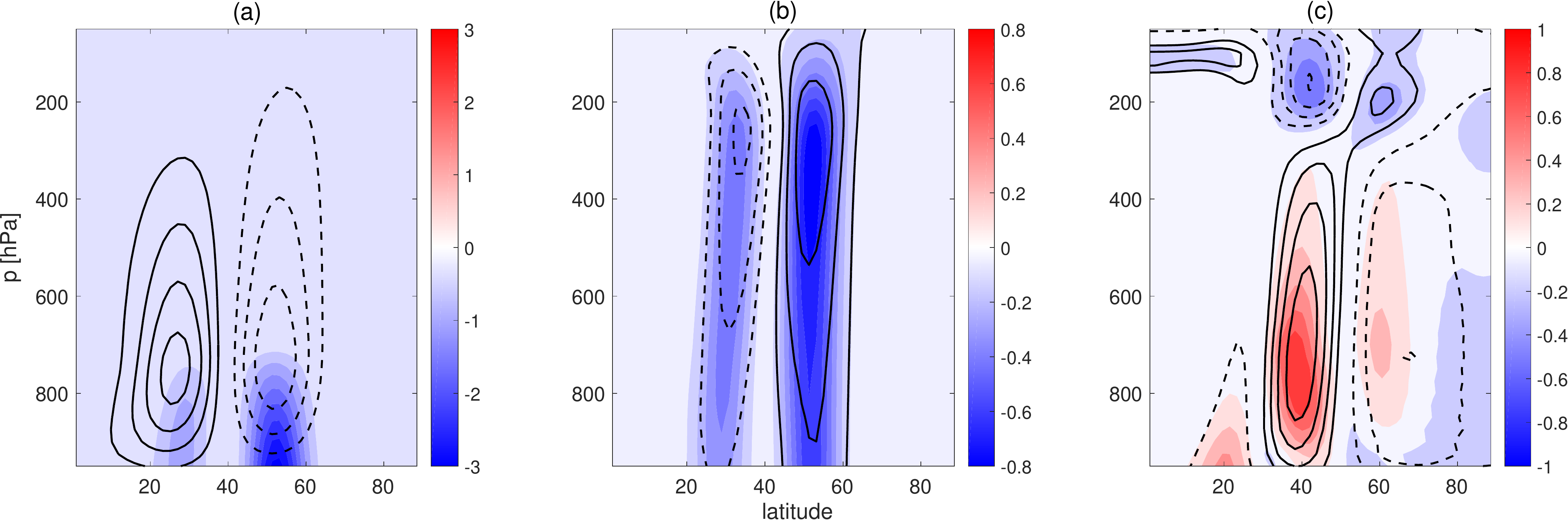}}
\caption{Effect of surface friction. (a) Shading shows $\{-k_v \, \overline{u}_\mathrm{AM}\}_u$~[day$^{-1}$] (see Eq.~(\ref{eq:Au}) for the definition of normalization $\{ \; \}_u$). Contourlines show the streamlines of the induced secondary circulation; Solid (dashed) lines indicate clockwise (counter-clockwise) circulation. (b) Shading shows the total tendency of $\overline{u}_\mathrm{AM}$ resulting from surface friction normalized according to Eq.~(\ref{eq:Au}). Unit is day$^{-1}$. Contourlines show $\overline{u}_\mathrm{AM}$; Solid (dashed) lines indicate westerly (easterly) wind. (c) Shading shows the total tendency of $\overline{T}_\mathrm{AM}$ resulting from surface friction normalized according to Eq.~(\ref{eq:AT}). Unit is day$^{-1}$. Contourlines show $\overline{T}_\mathrm{AM}$; Solid (dashed) lines indicate positive (negative) temperature.} 
\label{fig:uD}
\end{figure*}

Figure~\ref{fig:uD} shows how surface friction affects the annular mode zonal wind and temperature anomalies. Shading in Figure~\ref{fig:uD}(a) depicts the normalized $-k_v \overline{u}_\mathrm{AM}$ where $\overline{u}_\mathrm{AM}=\langle \overline{u} \rangle_\mathrm{EXP1}-\langle \overline{u} \rangle_\mathrm{CTL}$ is used. Hereafter, the normalization of any tendency $A$ is conducted as
\begin{eqnarray}
\{A\}_u&=& \frac{A \; \overline{u}_\mathrm{AM}}{\llbracket \overline{u}_\mathrm{AM} \overline{u}_\mathrm{AM} \rrbracket}, \label{eq:Au}\\
\{A\}_T&=& \frac{A \; \overline{T}_\mathrm{AM}}{\llbracket \overline{T}_\mathrm{AM} \overline{T}_\mathrm{AM} \rrbracket}. \label{eq:AT}
\end{eqnarray}     
Given this normalization, any red (blue) shading means strengthening (weakening) of the annular mode anomaly. 

As expected, the direct effect of surface friction is to weaken $\overline{u}_\mathrm{AM}$ at low levels. Using the solver for Eliassen problem, the secondary circulation induced by this forcing is calculated and found to consist of a pair of counter-rotating circulations with descending branches around $40^\mathrm{o}$ (Figure~\ref{fig:uD}(a)). North of $40^\mathrm{o}$, the poleward (equatorward) velocity at the lower  (upper) levels leads to anomalous westerly (easterly) tendency mostly through the term $+f \overline{v}$ in Eq.~(\ref{eq:u}), which strengthens (weakens) $\overline{u}_\mathrm{AM}$ at lower (upper) levels. The effect south of $40^\mathrm{o}$ can be similarly understood. Combined with the direct effect of friction, the total contribution of surface friction is weakening of the $\overline{u}_\mathrm{AM}$ anomaly throughout the troposphere (Figure~\ref{fig:uD}(b)). Surface friction also affects $\overline{T}_\mathrm{AM}$ via the secondary circulation: adiabatic warming from the descending flow around $40^\mathrm{o}$ strengthens $\overline{T}_\mathrm{AM}$ and increases the low-level baroclinicity (Figure~\ref{fig:uD}(c)). The effect of surface friction on $\overline{u}_\mathrm{BTRO}$ and $\overline{T}_\mathrm{BTRO}$ is qualitatively similar (Figure~\ref{fig:uDB}).

\subsection{Thermal Radiation} \label{subsec:rad}
\begin{figure*}[t]
\centerline{\includegraphics[width=1\textwidth]{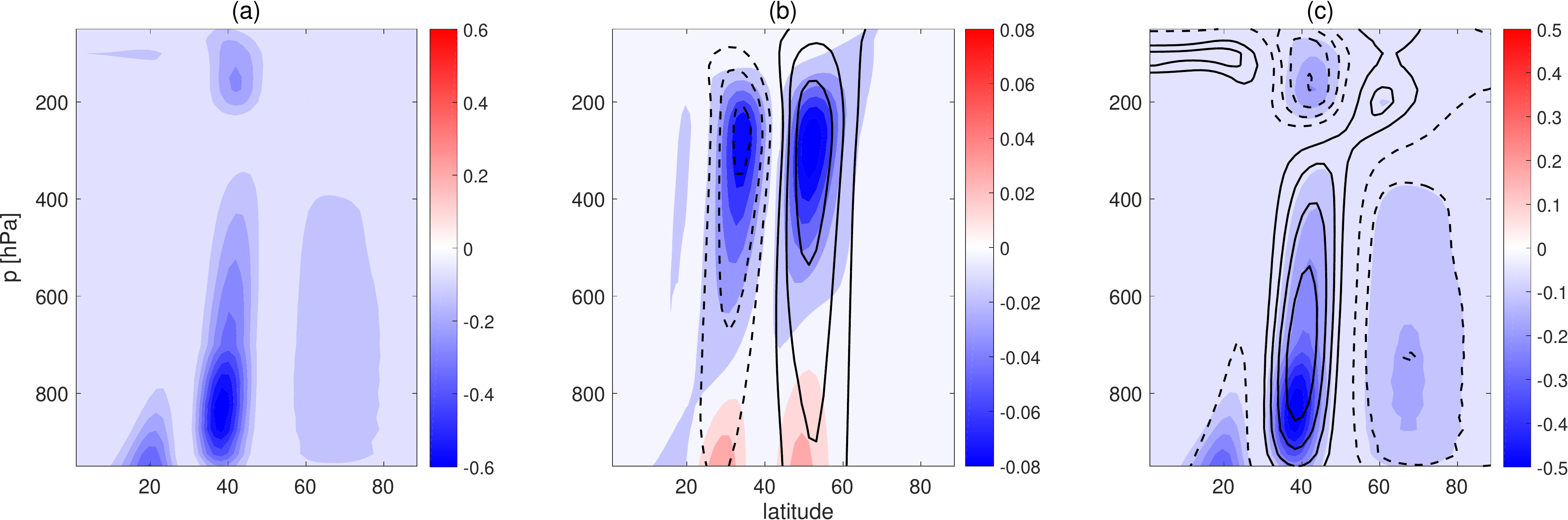}}
\caption{Effect of thermal radiation. (a) Shading shows $\{-k_T \, \overline{T}_\mathrm{AM}\}_T$ (see Eq.~(\ref{eq:AT}) for the definition of normalization $\{ \; \}_T$). The induced secondary circulation is weak and because the streamlines have the same intervals as in other figures, they do not show. (b) Shading shows the total tendency of $\overline{u}_\mathrm{AM}$ resulting from thermal radiation normalized according to Eq.~(\ref{eq:Au}). (c) Shading shows the total tendency of $\overline{T}_\mathrm{AM}$ resulting from thermal radiation normalized according to Eq.~(\ref{eq:AT}). Shadings in all panels have unit of day$^{-1}$. See the caption of Figure~\ref{fig:uD} for more details.} 
\label{fig:TD}
\end{figure*}

Figure~\ref{fig:TD} shows how thermal radiation affects the annular mode zonal wind and temperature anomalies. Shading in Figure~\ref{fig:TD}(a) depicts the normalized $-k_T \overline{T}_\mathrm{AM}$ where $\overline{T}_\mathrm{AM}=\langle \overline{T} \rangle_\mathrm{EXP1}-\langle \overline{T} \rangle_\mathrm{CTL}$ is used. As expected, the direct effect of thermal radiation is to weaken $\overline{T}_\mathrm{AM}$. The secondary circulation induced by this forcing is calculated using the solver for Eliassen problem. The induced circulation is weaker compared to the one in response to surface friction, but the resulting tendency on $\overline{u}_\mathrm{AM}$ has a similar pattern: weakening of the anomaly at the upper levels and strengthening at the lower levels (Figure~\ref{fig:TD}(b)). The total effect of thermal radiation is weakening of the $\overline{T}_\mathrm{AM}$ anomaly. Overall, the magnitude of the contribution of thermal radiation to the tendency of zonal wind (temperature) is a factor of 10 (2) smaller than the contribution of surface friction. The effect of thermal radiation on $\overline{T}_\mathrm{BTRO}$ and $\overline{T}_\mathrm{BTRO}$ has a similar pattern but is significantly weaker as expected (Figure~\ref{fig:TDB}).           

\subsection{Eddy Momentum Flux} \label{subsec:mom}
\begin{figure*}[t]
\centerline{\includegraphics[width=1\textwidth]{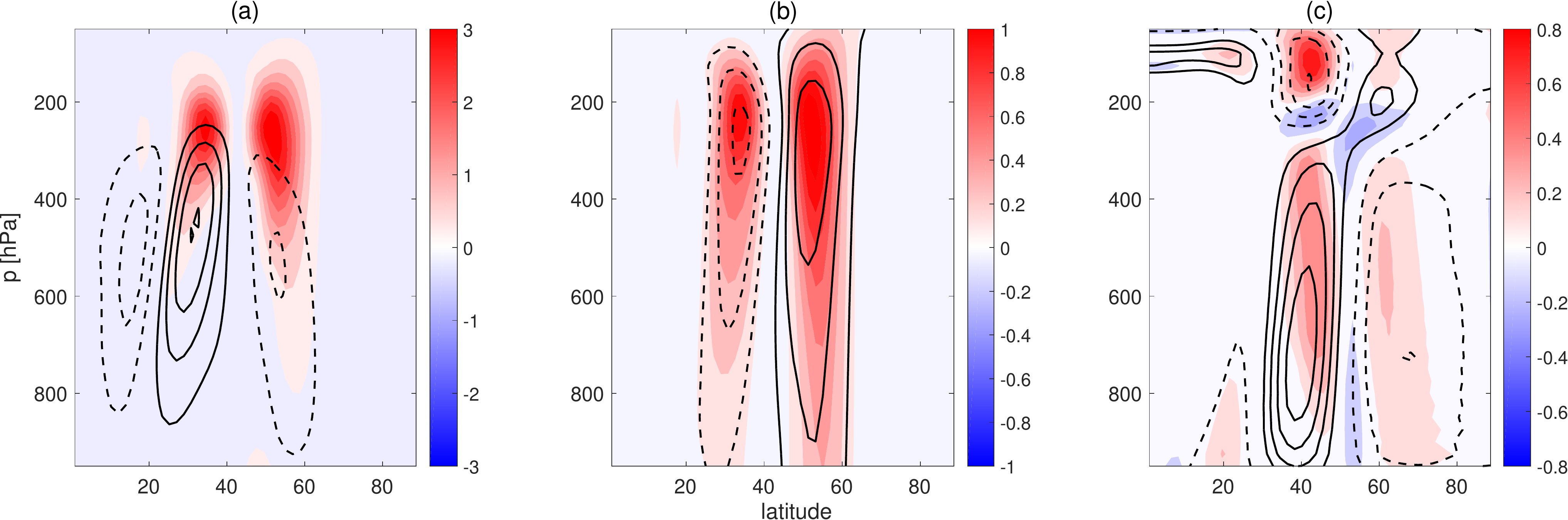}}
\caption{Effect of meridional eddy momentum flux divergence. (a) Shading shows $\{-\partial \overline{u'v'}_\mathrm{AM}/\partial y \}_u$ (calculations are conducted in spherical coordinates and $\partial /\partial y$ is used for convenience). (b) Shading shows the total tendency of $\overline{u}_\mathrm{AM}$ normalized according to Eq.~(\ref{eq:Au}). (c) Shading shows the total tendency of $\overline{T}_\mathrm{AM}$ normalized according to Eq.~(\ref{eq:AT}). Shadings in all panels have unit of day$^{-1}$. See the caption of Figure~\ref{fig:uD} for more details.} 
\label{fig:uv}
\end{figure*}

Figure~\ref{fig:uv} shows how meridional eddy momentum flux divergence affects the annular mode zonal wind and temperature anomalies. Shading in Figure~\ref{fig:uv}(a) depicts the normalized $-\partial \overline{u'v'}_\mathrm{AM} / \partial y$ where $\overline{u'v'}_\mathrm{AM}=\langle \overline{u'v'} \rangle_\mathrm{EXP1}-\langle \overline{u'v'} \rangle_\mathrm{CTL}$ is used. The direct effect of the meridional eddy momentum flux is to strengthen $\overline{u}_\mathrm{AM}$ at the upper levels. The induced secondary circulation (shown in Figure~\ref{fig:uv}(a)) transfers some of the anomalous westerly (easterly) momentum to the low levels north (south) of $40^\mathrm{o}$, leading to strengthening of $\overline{u}_\mathrm{AM}$ throughout the atmosphere (Figure~\ref{fig:uv}(b)). The induced secondary circulation also strengthens $\overline{T}_\mathrm{AM}$ particularly around $40^\mathrm{o}$ above $700$~hPa. The direct and total effects of vertical eddy momentum flux divergence have been also calculated but found to be very weak (not shown). 
  
The effect of meridional eddy momentum flux divergence in response to only the barotropic component of the annular mode is quite different (as discussed in section~\ref{sec:forced}\ref{sec:forced_BT}) and leads to weakening of both $\overline{u}_\mathrm{AM}$ and $\overline{T}_\mathrm{AM}$, see Figure~\ref{fig:uvB}.  

\subsection{Eddy Heat Flux} \label{subsec:heat}
\begin{figure*}[t]
\centerline{\includegraphics[width=1\textwidth]{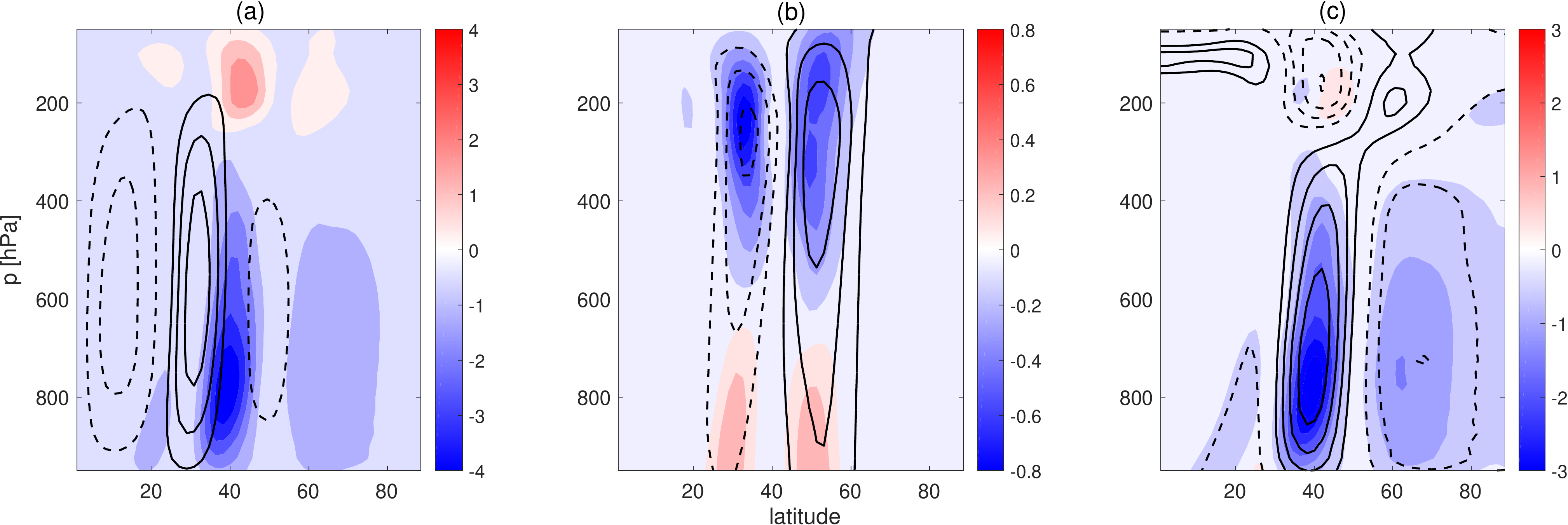}}
\caption{Effect of meridional eddy heat flux divergence. (a) Shading shows $\{-\partial \overline{v'T'}_\mathrm{AM}/\partial y \}_T$ (calculations are conducted in spherical coordinates and $\partial /\partial y$ is used for convenience). (b) Shading shows the total tendency of $\overline{u}_\mathrm{AM}$ normalized according to Eq.~(\ref{eq:Au}). (c) Shading shows the total tendency of $\overline{T}_\mathrm{AM}$ normalized according to Eq.~(\ref{eq:AT}). Shadings in all panels have unit of day$^{-1}$. See the caption of Figure~\ref{fig:uD} for more details.}  
\label{fig:vT}
\end{figure*}

Figure~\ref{fig:vT} shows how meridional eddy heat flux divergence affects the annular mode zonal wind and temperature anomalies. Shading in Figure~\ref{fig:vT}(a) depicts the normalized $-\partial \overline{v'T'}_\mathrm{AM} / \partial y$ where $\overline{v'T'}_\mathrm{AM}=\langle \overline{v'T'} \rangle_\mathrm{EXP1}-\langle \overline{v'T'} \rangle_\mathrm{CTL}$ is used. The induced secondary circulation has equatorward (poleward) velocity at upper (lower) levels between $40^\mathrm{o}-60^\mathrm{o}$ (and the opposite between $40^\mathrm{o}-60^\mathrm{o}$) which mainly via the $f \overline{v}$ term, leads to the weakening (strengthening) of $\overline{u}_\mathrm{AM}$ at the upper (lower) levels as shown in Figure~\ref{fig:vT}(b). 

\begin{figure*}[]
\centerline{\includegraphics[width=1\textwidth]{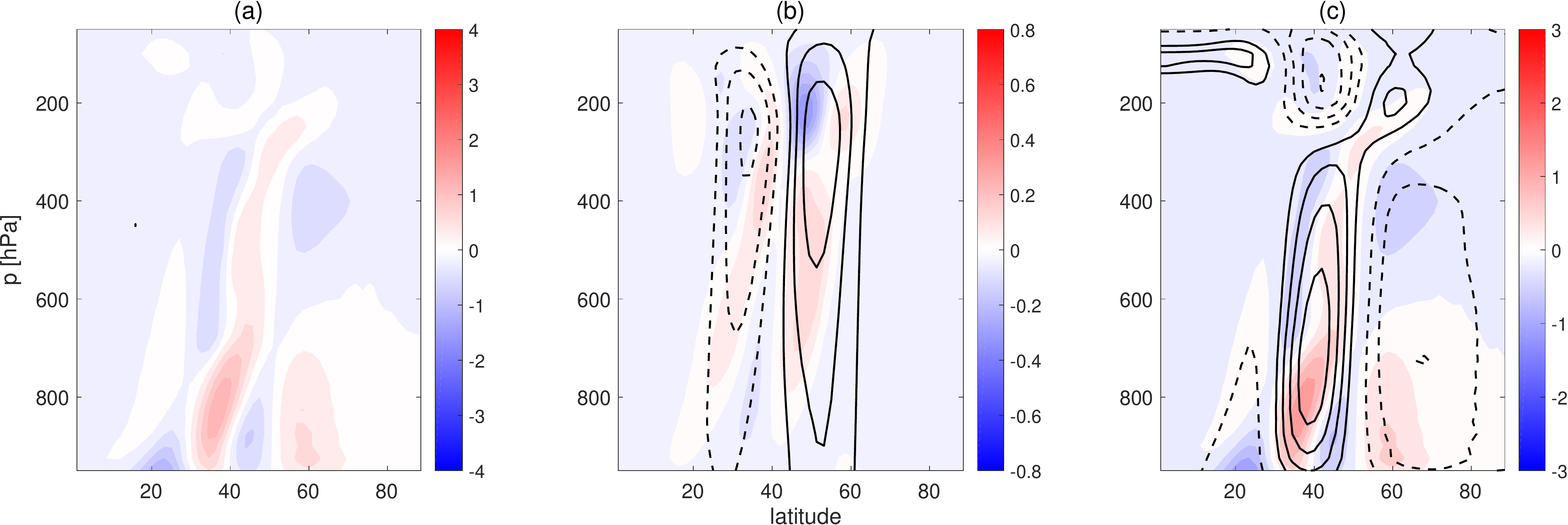}}
\caption{Same as Figure~\ref{fig:vT} but for vertical eddy heat flux divergence. The induced secondary circulation is weak and because the streamlines have the same intervals as in other figures, they do not show.} 
\label{fig:wT}
\end{figure*}

The total effect of meridional eddy heat flux on $\overline{T}_\mathrm{AM}$ is to weaken the anomaly. The direct and total effects of vertical eddy heat flux divergence have been calculated similarly and shown in Figure~\ref{fig:wT}. Although the magnitude of the total tendencies are weaker than those of the meridional eddy heat flux, there is noticeable influence on the low-level baroclinicity which is further discussed in section~\ref{sec:res}.

The effect of meridional eddy heat flux divergence in response to only the barotropic component of the annular mode is quite different and leads to strengthening (weakening) of $\overline{u}_\mathrm{AM}$ at the upper levels (lower levels) and strengthening of $\overline{T}_\mathrm{AM}$,  see Figure~\ref{fig:vTB}. 

\begin{figure*}[t]
\centerline{\includegraphics[width=1\textwidth]{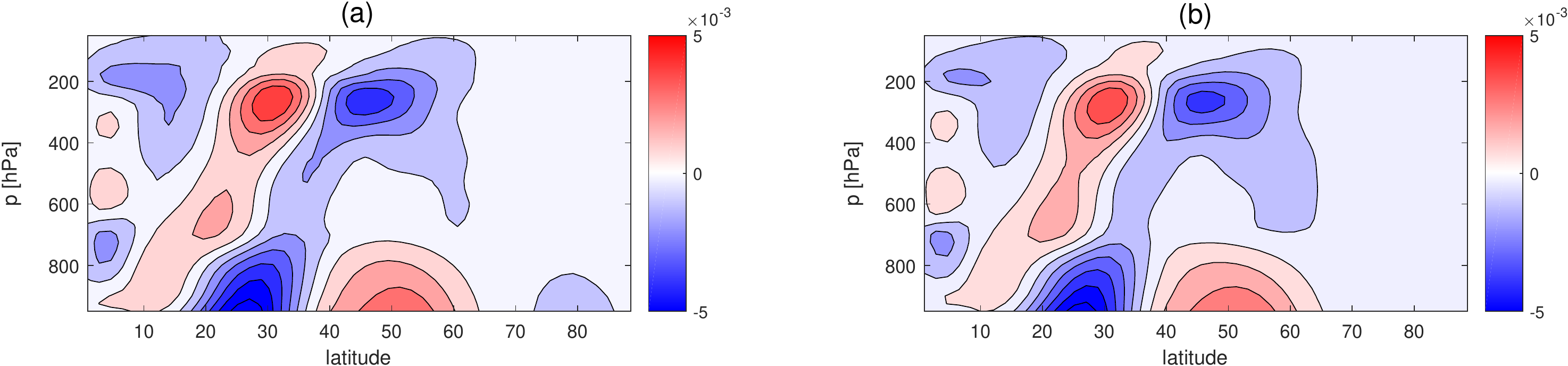}}
\caption{The meridional wind of the induced secondary circulation: (a) $\overline{v}_\mathrm{AM}=\langle \overline{v} \rangle_\mathrm{EXP1} - \langle \overline{v} \rangle_\mathrm{CTL}$. (b) Sum of all induced meridional velocities calculated individually from solving the Eliassen problem. Unit is m~s$^{-1}$.}  
\label{fig:V}
\end{figure*}

The direct and total effects of advection of $\overline{u}_\mathrm{AM}$ and $\overline{T}_\mathrm{AM}$ by meridional circulation $(\overline{V},\overline{\Omega})$ and $\overline{\boldsymbol{f}}_\mathrm{EXP1}$ have been also calculated similarly (not shown). As a test for the accuracy of the calculations described above, we compare in Figure~\ref{fig:V} the meridional wind of the annular mode's secondary circulation anomaly $\overline{v}_\mathrm{AM}=\langle \overline{v} \rangle_\mathrm{EXP1} - \langle \overline{v} \rangle_\mathrm{CTL}$ with the sum of all the individually calculated induced meridional winds that are calculated from solving the Eliassen problem. The agreement in pattern and magnitude of the two meridional velocities indicates the accuracy of the numerical procedure used to solve the Eliassen problem.    

\section{Annular Mode Dynamics} \label{sec:res}

\begin{figure*}[t]
\centerline{\includegraphics[width=1\textwidth]{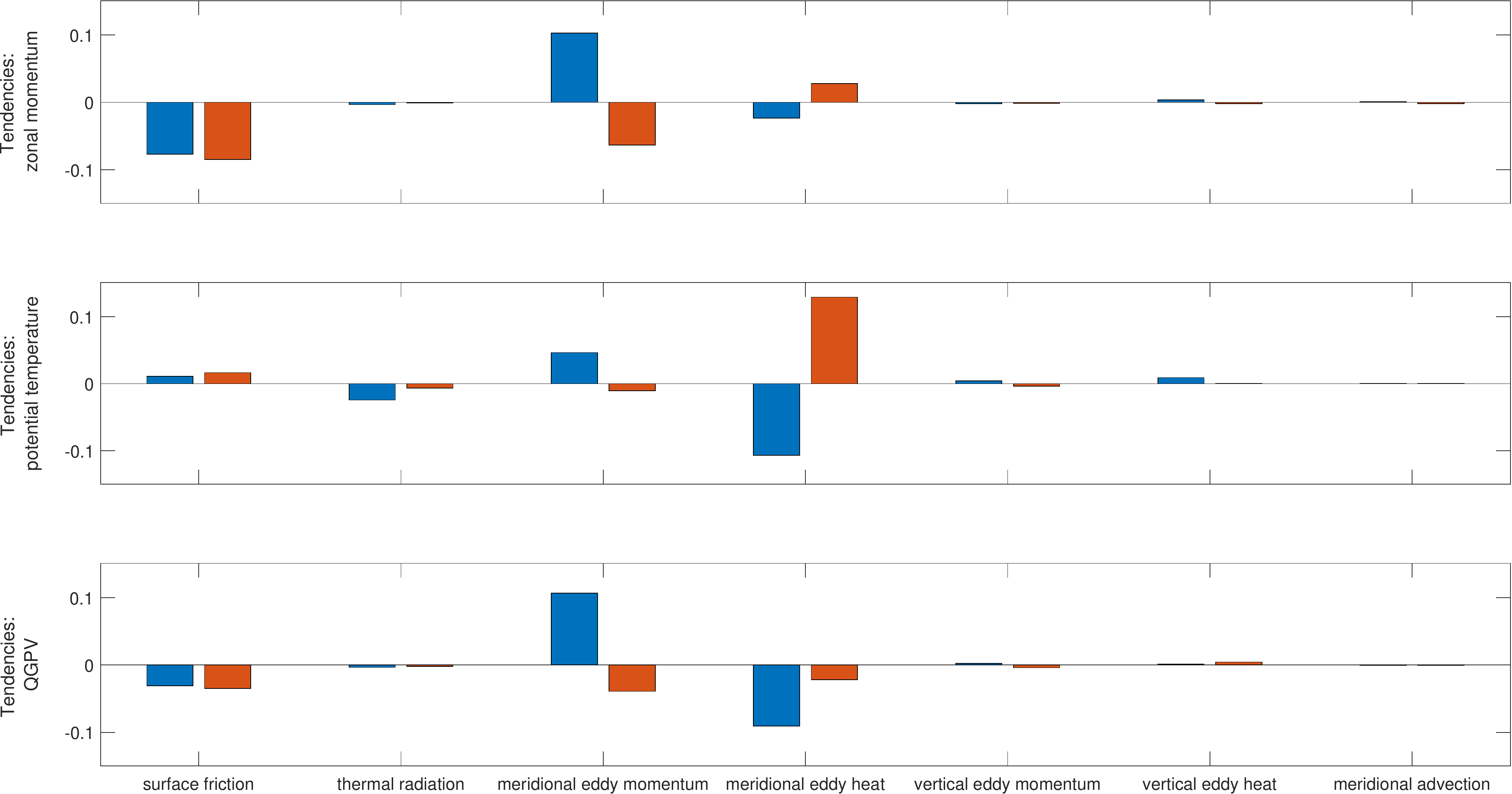}}
\caption{Contribution of different processes to the tendency budget of $\overline{u}_\mathrm{AM}$ (top panel), $\overline{\theta}_\mathrm{AM}$ (middle panel), and quasi-geostrophic potential vorticity QGPV (bottom panel).  The vertical axes have the unit of day$^{-1}$ and show, from top panel to bottom panel, $\llbracket \{ A \}_u \rrbracket$, $\llbracket \{ A \}_T \rrbracket$, and $\llbracket \{ A \}_{QGPV} \rrbracket$ for tendency $A$, respectively. Blue (red) bars indicate the contribution to the full (barotropic-only component of the) annular mode anomaly.}  
\label{fig:balance}
\end{figure*}

The blue bars in Figure~\ref{fig:balance} show the contribution of area-weighted domain-averaged normalized total tendency of each process, calculated in the previous section, to the zonal momentum balance, potential temperature balance, and quasi-geostrophic potential vorticity (QGPV) balance of the full annular mode anomaly. For the zonal momentum balance, as expected, surface friction causes a negative tendency, while meridional eddy momentum flux is the largest (and only non-negligible) source of positive tendency. The only other non-negligible contribution is a small negative tendency from the meridional heat flux, mainly from weakening of the upper-level wind anomaly (Figure~\ref{fig:vT}(b)). The contribution from other terms is negligible, not only because of cancellations between positive and negative tendencies in the domain, but also (and in fact, mainly) because the magnitudes are small. 

In the potential temperature balance, as expected, thermal radiation and meridional eddy heat flux cause negative tendencies while meridional eddy momentum flux, surface friction, and vertical eddy heat flux cause positive tendencies. Although the positive tendency from surface friction is small, its importance should be further emphasized. The large positive tendency from meridional eddy momentum flux is mostly dominated by the positive tendency in the stratosphere and to some extent, between $300-800$~hPa around $40^\mathrm{o}$ (Figure~\ref{fig:uv}(c)). The positive tendency from surface friction on the other hand, is concentrated below $600$~hPa around $40^\mathrm{o}$ (Figure~\ref{fig:uD}(c)) and reinforces the anomalous low-level baroclinicity, which increases eddy generation and leads to the positive eddy forcing from meridional eddy momentum flux at the upper levels \citep{robinson2000baroclinic,lorenz2001eddy}. The tendency from the vertical eddy heat flux is also mostly concentrated below $700$~hPa between $20^\mathrm{o}$ and $60^\mathrm{o}$ although there is a complex pattern of positive/negative tendencies (Figure~\ref{fig:wT}(c)).          

The QGPV tendencies describe the balance of the annular mode anomaly as a system, showing that overall, the anomaly persists because the meridional eddy momentum flux balances the negative tendencies from surface friction and meridional eddy heat flux. The sum of all tendencies yields a decay rate of $-0.016$~day$^{-1}$, or a $\sim 62$~day timescale. Note that because we are looking at the steady-state balance of EXP1, the forcing $\overline{\boldsymbol{f}}_\mathrm{EXP1}$, which is calculated to have a total tendency of $+0.018$~day$^{-1}$,  is indeed balancing the net tendency from physical processes. The residual of these two tendencies, $\sim 0.002$~day$^{-1}$ is a measure of uncertainty in the calculated balance.          

Without the contributions from the eddy fluxes, the sum of all other tendencies yields a decay rate of $-0.035$~day$^{-1}$, or a $\sim 28$~day timescale, indicating that the \emph{positive} feedbacks from eddy fluxes that are in response to the full annular mode anomaly are {increasing} the annular mode persistence by a factor of $\sim 2.3$. 

The red bars in Figure~\ref{fig:balance} show the contribution of area-weighted domain-averaged normalized total tendency of each process for the barotropic-only component of the annular mode. In the zonal momentum and potential temperature balances, the effect of all processes except for meridional eddy fluxes are qualitatively the same as for the full annular mode anomaly. The signs of the tendencies from meridional eddy fluxes are opposite, consistent with the discussions and results in section~\ref{sec:forced} and Appendix~C. In the QGPV balance, the tendencies from surface friction and meridional eddy momentum flux are negative, consistent with the results in the first two panels. The contribution from the meridional eddy heat flux is also negative in spite of the positive tendencies in the first two panels. Further examination shows that this tendency in QGPV is dominated by the contribution to the zonal momentum balance, which once converted into QGPV and projected onto the annular mode anomaly, leads to an overall negative tendency. 

The sum of all tendencies for the barotropic-only component yields a decay rate of $-0.098$~day$^{-1}$, or a $\sim 10$~day timescale (taking the tendency of $\overline{\boldsymbol{f}}_\mathrm{EXP2}$ into account, the residual tendency is $-0.010$~day$^{-1}$). Comparing this timescale with that of the balance without eddy fluxes, where the timescale is $\sim 28$~day, these results suggest that there is a \emph{negative} eddy-jet feedback between the barotropic component of the annular mode anomalies and the eddies, which \emph{decreases} the persistence of the annular mode by a factor of $\sim 2.8$. 
                           

\section{Discussion and Summary} \label{sec:sum}
\begin{figure*}[t]
\centerline{\includegraphics[width=0.75\textwidth]{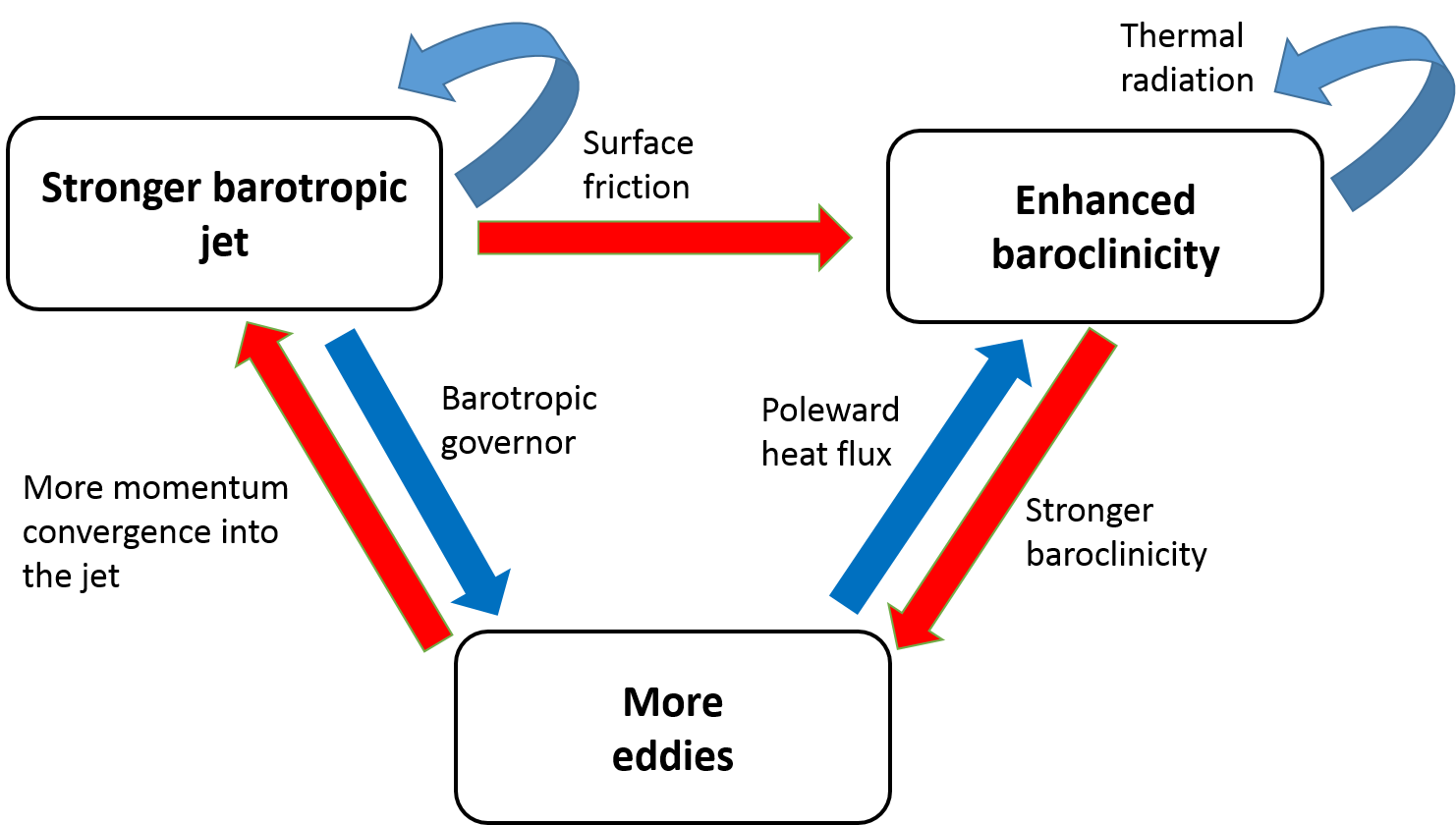}}
\caption{Schematic of the annular mode dynamics in the idealized atmosphere, focused on the poleward half of the annular mode anomaly ($40^\mathrm{o}-60^\mathrm{o}$) for convenience. Red (blue) arrows indicate positive (negative) influence.}  
\label{fig:schem}
\end{figure*}

The dynamics of the annular mode at the low-frequency ($\sim$ steady-state) limit that emerges from the above analysis is shown schematically in Figure~\ref{fig:schem}. Focusing on the poleward half of the annular mode anomaly ($40^\mathrm{o}-60^\mathrm{o}$) for convenience,  an anomalous increase in the eddies leads to a stronger barotropic jet, which in turn suppresses the generation of the eddies because of the barotropic governor effect. The stronger barotropic jet increases the low-level baroclinicity, which in turn, intensifies eddy generation. However, the increase in eddies reduces the baroclinicity through enhanced poleward heat flux. Note that through the negative influence of the barotropic governor effect on eddy generation and the negative influence of the poleward heat flux on the low-level baroclinicity (the two straight blue arrows in Figure~\ref{fig:schem}), the stronger barotropic jet has a further positive influence on the low-level baroclinicity (see Figure~\ref{fig:vTB}(c)). However, the net effect of this influence is reduction in eddies and a negative eddy forcing (see Figure~\ref{fig:uvB}(b)).
   
In summary, using the LRF framework of \citet{hassanzadeh2016linear} and the same idealized setup as \citet{ma2017quantifying}, in this paper 
\begin{enumerate}
\item The contribution of different physical processes (eddy momentum flux, eddy heat flux, surface friction, thermal radiation, and meridional advection) to the annular mode dynamics is quantified (sections~\ref{sec:quant} and \ref{sec:res}). This was made possible by first quantifying how eddy fluxes change in response to the annular mode anomaly (section~\ref{sec:forced}).     
\item It is shown in section~\ref{sec:res} that the eddies modified by the full (barotropic+baroclinic) annular mode anomaly exert a \emph{positive} feedback and \emph{increase} the annular mode persistence by a factor of $\sim 2.3$. It is further shown that the eddies modified by only the barotropic component of the annular mode anomaly exert a \emph{negative} feedback and \emph{decrease} the annular mode persistence by a factor of $\sim 2.8$.   
\item It is shown in section~\ref{sec:forced} that how eddy fluxes change in response to the full annular mode anomaly is drastically different from their response to the barotropic-only component of the anomaly and that a positive eddy-jet feedback exists only for the former. Results suggest that the baroclinic component of the annular mode anomaly, i.e., increased low-level baroclinicity, is essential for the persistence of the annular mode, consistent with the work of \citet{robinson2000baroclinic} and \citet{lorenz2001eddy}.
\end{enumerate}

Further analyses in future studies are needed to understand the implications and relevance of these results beyond the idealized atmosphere studied here. In this study, we aimed to examine the annular mode dynamics internal to the troposphere in a minimal primitive-equation model of the annular mode. Thus we focused on an idealized GCM setup that is at equinox and does not have a dynamically active stratosphere or topography. The extratropical circulation in this setup is dominated by EOF1, which explains $51\%$ of the variance compared to the $17\%$ variance explained by EOF2. Recently \citet{sheshadri2017propagating} have shown that if the setup is changed to mimic winter/summer seasons, then in the winter hemisphere, the EOF1 and EOF2 have comparable explained variances ($40\%$ and $29\%$), resulting in a circulation that is dominated by a propagating annular mode in which EOF1 and EOF2 are correlated at low-frequency and feedback into each other. Repeating the analysis presented here for such setup can provide insight into the dynamics of propagating annular modes. 

Furthermore, including topography and other zonal asymmetries to examine the feedback from planetary waves \citep{riviere2016short} and adding an active stratosphere to investigate the effect of stratospheric variability on the annular mode persistence \citep{byrne2017nonstationarity,byrne2018seasonal} can help with developing a deeper understanding of the annular mode dynamics in the NH and SH. 

Finally, the barotropic (baroclinic) annular mode is sometimes considered as variability of eddy momentum (heat) flux without much connection with the eddy heat (momentum) flux, suggesting that the two annular modes are decoupled \citep{thompson2014barotropic}. The results presented here show the essential role of eddy heat flux and baroclinicity in the dynamics of the barotropic annular mode in the low-frequency limit, suggesting that the two annular modes are potentially coupled at such timescales, consistent with the recent findings of \citet{boljka2018coupling}. A deeper understanding of this coupling is needed for a complete model of the extratropical low-frequency variability. 

The main complication in repeating the analysis conducted here for different GCM setups is that the LRF has to be recomputed for each setup using the Green's function method. This method requires many forced simulations of the GCM, which can be computationally cumbersome, but otherwise the Green's function method has been successfully applied to compute the LRFs for a variety of models \citep{kuang2010linear,kuang2012weakly,hassanzadeh2016linear,liu2018sensitivity,khodkar2018reduced}. The FDT, which can provide the LRF from the data of one long, unforced simulation, is an attractive alternative to the Green's function method; however, the application of FDT to GCMs has produced mixed results \citep{ring08,fuchs2015exploration,lutsko2015applying,hassanzadeh2016linear2}. \citet{khodkar2018data} have recently introduced a new FDT framework that has shown a promising performance for a model of turbulent convection; if this framework works comparably well for GCMs, then repeating the analysis of this paper for other model setups will be significantly facilitated.           
              

\acknowledgments
We thank David Lorenz, Ding Ma and Aditi Sheshadri for fruitful discussions. This work was supported by NSF grant AGS-1552385 and NASA grant 80NSSC17K0266. The simulations were run on the Odyssey cluster supported by the FAS Division of Science, Research Computing Group at Harvard University. The Rice University Center for Research Computing and XSEDE Stampede2 (via allocation ATM170020) provided additional computational resources.  


\setcounter{figure}{0}    
\appendix[A]
\appendixtitle{Parameterizations in the idealized GCM}
\label{app:A}
The parameterizations of thermal radiation as Newtonian relaxation $-k_T (T-T_{eq})$ and planetary boundary layer as Rayleigh drag $-k_v {\boldsymbol{v}}$ are identical to those described in \citet{held1994proposal}: The equilibrium temperature profile is  
\begin{eqnarray}
\!\!\! T_{eq}(\phi,p)={\mathrm{max}} \bigg\{200, T_{trop} \bigg\},
\label{eq:Teq}
\end{eqnarray}
where 
\begin{eqnarray}
T_{trop} = \left[ 315 - \Delta T_y \sin^2{\phi} -
\Delta \theta_z \log\left[{\frac{p}{p_o}}\right]\cos^2{\phi}  \right] \left[\frac{p}{p_o}\right]^2
\end{eqnarray}
and $\Delta T_y=60$~K, $\Delta \theta_z=10$~K, and $p_o=1000$~hPa.
The Newtonian relaxation rate is
\begin{eqnarray}
k_T(\phi,p)=k_a+(k_s-k_a) \; {\mathrm{max}} \bigg\{0,\frac{\sigma-\sigma_b}{1-\sigma_b} \bigg\}  \cos^4{\phi},
\label{eq:kT}
\end{eqnarray}
where $k_a=1/40$~day$^{-1}$, $k_s=1/4$~day$^{-1}$, and $\sigma_b=0.7$.
The Rayleigh drag damping rate is
\begin{eqnarray}
k_v(\phi,p)=k_f \; {\mathrm{max}} \bigg\{0,\frac{\sigma-\sigma_b}{1-\sigma_b} \bigg\},
\label{eq:kv}
\end{eqnarray}
where $k_f=1$~day$^{-1}$.

\appendix[B]
\appendixtitle{GCM solver for Eliassen problem}
\label{app:B}
Several changes have to be made to the GCM to turn it into an axisymmetric solver for the Eliassen problem described in section~\ref{sec:quant} and in \citet[][appendix~A]{ring08}. First
\begin{enumerate}
\item The model is initialized with a completely zonally symmetric $(\overline{U}(\phi,\sigma),\overline{\Theta}(\phi,\sigma))$.
\item A small zonally asymmetric perturbation that is added in the GCM to hasten the baroclinic instability is removed.
\item The Rayleigh drag and Newtonian cooling of the Held-Suarez physics are disabled.
\item The optional corrections to conserve mass and energy are disabled.
\item The coefficient of the Robert--Asselin time filter is set to $0$. 
\item For time integration, the fully backward scheme is used (by choosing $\alpha=1$) to help with damping/slowing down the gravity waves.   
\end{enumerate}
For items 4-6, see the description of the model and algorithms of the GFDL spectral dynamical core\footnote{\url{https://www.gfdl.noaa.gov/wp-content/uploads/files/user_files/pjp/spectral_core.pdf}}. 

With these changes, for $\overline{U}$ and $\overline{\Theta}$ that are in exact gradient-wind balance, the model can be integrated for thousands of days without any change in zonal wind and temperature. However, the $\overline{U}$ and $\overline{\Theta}$ obtained from the climatology of the CTL are not in exact gradient-wind balance which leads to generation of waves and secondary circulations. To resolve this issue, we calculated the tendency of the meridional wind $\partial \overline{v}/\partial t$ after one time step, and then from the meridional momentum equation, calculated the change in the zonal wind $\Delta \overline{U}$ needed to balance this tendency:
\begin{eqnarray}
\frac{\partial \overline{v}}{\partial t}=(\overline{U}+\Delta \overline{U})\left(f+\frac{\overline{U}+\Delta \overline{U}}{a} \tan{\phi}   \right). 
\end{eqnarray}      
$\Delta \overline{U}$ is computed from solving a quadratic equation. Now initializing the model with $(\overline{U}+\Delta \overline{U},\overline{\Theta})$ leads to a balanced system that can be integrated for thousands of days without developing any secondary circulation. Note that the issue described above is not a pitfall of our approach to solving the balance Eliassen problem. An exact gradient-wind balance is the key assumption of the Eliassen problem, and in other approaches, for example the one described in appendix~A of \citet{ring08}, exact gradient-wind balance is assumed and linearized gradient-wind balance is enforced during the formulation.    

To compute the induced secondary circulation and total tendency from a forcing, e.g., $-k_v \overline{u}_\mathrm{AM}$, this forcing is added to both hemispheres of this model and the mean $\overline{v}$, $\partial \overline{u}/\partial t$, and $\partial \overline{T}/\partial t$ are computed between days $50$ and $100$ (the results are not sensitive to the exact time period used). These are, respectively, the induced meridional velocity of the secondary circulation, and the total tendency of zonal wind and temperature resulting from this forcing.       

\begin{figure*}[t]
\centerline{\includegraphics[width=1\textwidth]{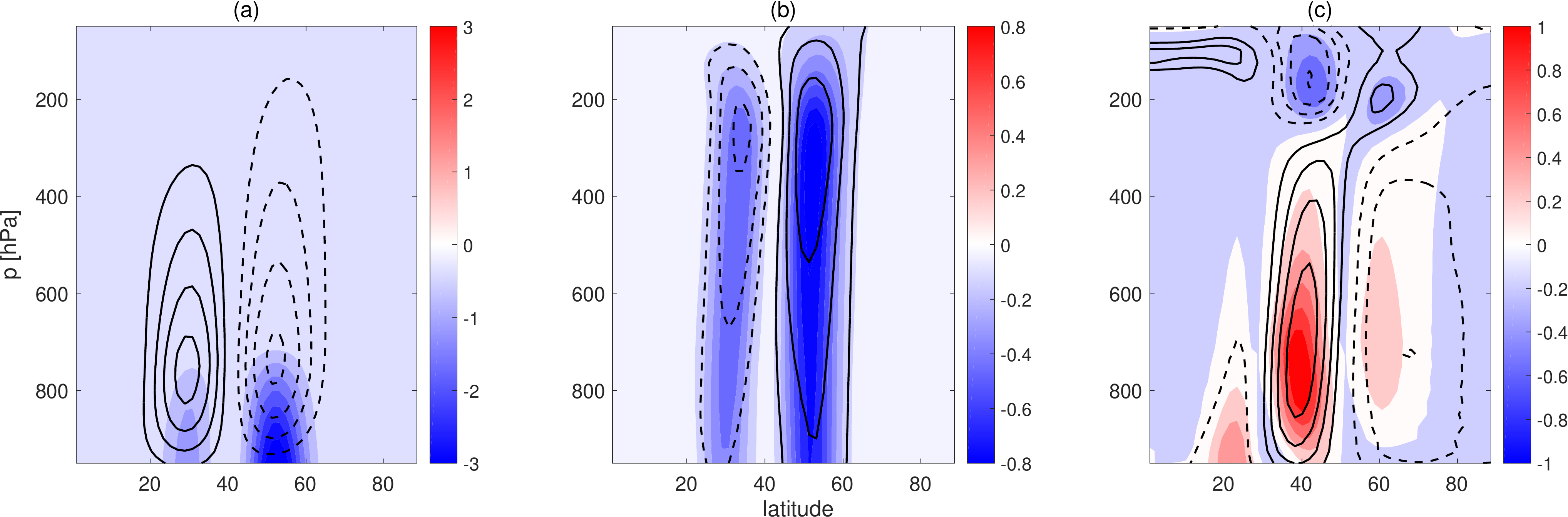}}
\caption{Same as Figure~\ref{fig:uD} but for the effect of surface friction on the barotropic component of the annular mode: $-k_v \, \overline{u}_\mathrm{BTRO}$ where $\overline{u}_\mathrm{BTRO}=\langle \overline{u} \rangle_\mathrm{EXP2}-\langle \overline{u} \rangle_\mathrm{CTL}$.} 
\label{fig:uDB}
\end{figure*}

\begin{figure*}[t]
\centerline{\includegraphics[width=1\textwidth]{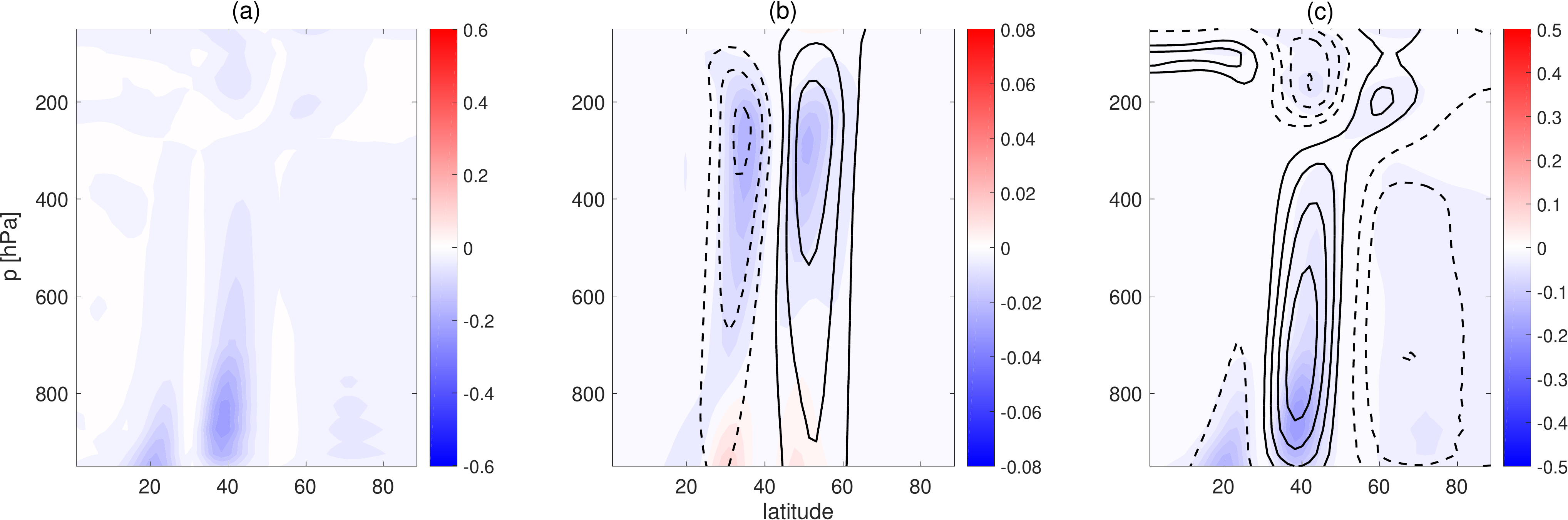}}
\caption{Same as Figure~\ref{fig:TD} but for the effect of thermal radiation on the barotropic component of the annular mode: $-k_T \, \overline{T}_\mathrm{BTRO}$ where $\overline{T}_\mathrm{BTRO}=\langle \overline{T} \rangle_\mathrm{EXP2}-\langle \overline{T} \rangle_\mathrm{CTL}$.} 
\label{fig:TDB}
\end{figure*}

\begin{figure*}[t]
\centerline{\includegraphics[width=1\textwidth]{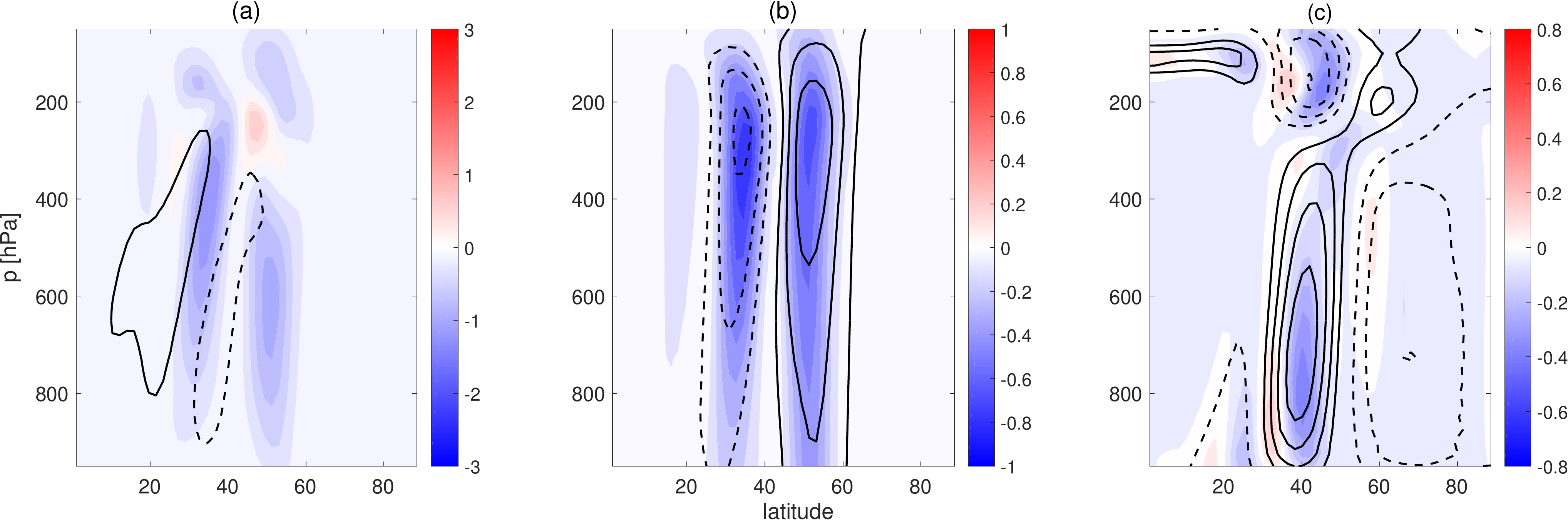}}
\caption{Same as Figure~\ref{fig:uv} but for the effect of meridional eddy momentum flux divergence in response to the barotropic component of the annular mode: $-\partial \overline{u'v'}_\mathrm{BTRO}/\partial y$ where $\overline{u'v'}_\mathrm{BTRO}=\langle \overline{u'v'} \rangle_\mathrm{EXP2}-\langle \overline{u'v'} \rangle_\mathrm{CTL}$.} 
\label{fig:uvB}
\end{figure*}

\begin{figure*}[t]
\centerline{\includegraphics[width=1\textwidth]{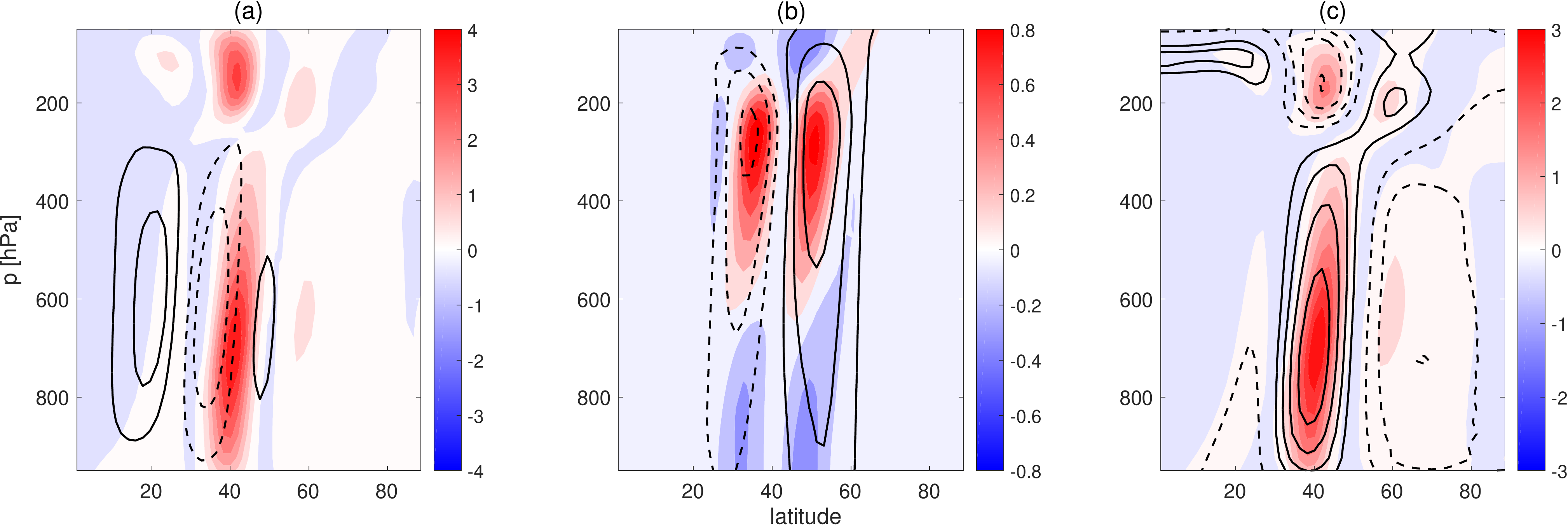}}
\caption{Same as Figure~\ref{fig:vT} but for the effect of meridional eddy heat flux divergence in response to the barotropic component of the annular mode: $-\partial \overline{v'T'}_\mathrm{BTRO}/\partial y$ where $\overline{v'T'}_\mathrm{BTRO}=\langle \overline{v'T'} \rangle_\mathrm{EXP2}-\langle \overline{v'T'} \rangle_\mathrm{CTL}$.} 
\label{fig:vTB}
\end{figure*}

\appendix[C]
\appendixtitle{Effect of different processes on $(\overline{u},\overline{T})_\mathrm{BTRO}$}
\label{app:C}
Four figures (\ref{fig:uDB}-\ref{fig:vTB}) that are the same as Figures~\ref{fig:uD}--\ref{fig:vT} but for the effect on the barotropic component of the annular mode, i.e., using the results of EXP2 instead of EXP1, are included in this appendix.  

%
%
%




\bibliographystyle{ametsoc2014}
\setlength{\bibsep}{2pt plus 0.75ex}
\bibliography{AODyn_v3}

\begin{thebibliography}{62}
\providecommand{\natexlab}[1]{#1}
\providecommand{\url}[1]{\texttt{#1}}
\renewcommand{\UrlFont}{\rmfamily}
\providecommand{\urlprefix}{URL }
\expandafter\ifx\csname urlstyle\endcsname\relax
  \providecommand{\doi}[1]{doi:\discretionary{}{}{}#1}\else
  \providecommand{\doi}{doi:\discretionary{}{}{}\begingroup
  \urlstyle{rm}\Url}\fi
\providecommand{\eprint}[2][]{\url{#2}}

\bibitem[{Balasubramanian and Garner(1997)Balasubramanian, and
  Garner}]{balasubramanian1997role}
Balasubramanian, G., and S.~T. Garner, 1997: The role of momentum fluxes in
  shaping the life cycle of a baroclinic wave. \textit{J.\ Atmos.\ Sci.},
  \textbf{54~(4)}, 510--533.

\bibitem[{Barnes et~al.(2010)Barnes, Hartmann, Frierson,, and
  Kidston}]{barnes2010effect}
Barnes, E.~A., D.~L. Hartmann, D.~M.~W. Frierson, and J.~Kidston, 2010: Effect
  of latitude on the persistence of eddy-driven jets. \textit{Geophys.\ Res.\
  Lett.}, \textbf{37~(11)}.

\bibitem[{Boljka et~al.(2018)Boljka, Shepherd,, and
  Blackburn}]{boljka2018coupling}
Boljka, L., T.~G. Shepherd, and M.~Blackburn, 2018: On the coupling between
  barotropic and baroclinic modes of extratropical atmospheric variability.
  \textit{J.\ Atmos.\ Sci.}, \textbf{75~(6)}, 1853--1871.

\bibitem[{Branstator(1995)}]{branstator1995organization}
Branstator, G., 1995: Organization of storm track anomalies by recurring
  low-frequency circulation anomalies. \textit{J.\ Atmos.\ Sci.},
  \textbf{52~(2)}, 207--226.

\bibitem[{Byrne and Shepherd(2018)Byrne, and Shepherd}]{byrne2018seasonal}
Byrne, N.~J., and T.~G. Shepherd, 2018: Seasonal persistence of circulation
  anomalies in the {Southern Hemisphere} stratosphere and its implications for
  the troposphere. \textit{J.\ Climate}, \textbf{31~(9)}, 3467--3483.

\bibitem[{Byrne et~al.(2016)Byrne, Shepherd, Woollings,, and
  Plumb}]{byrne2016annular}
Byrne, N.~J., T.~G. Shepherd, T.~Woollings, and R.~A. Plumb, 2016: Annular
  modes and apparent eddy feedbacks in the {Southern Hemisphere}.
  \textit{Geophys.\ Res.\ Lett.}, \textbf{43~(8)}, 3897--3902.

\bibitem[{Byrne et~al.(2017)Byrne, Shepherd, Woollings,, and
  Plumb}]{byrne2017nonstationarity}
Byrne, N.~J., T.~G. Shepherd, T.~Woollings, and R.~A. Plumb, 2017:
  Nonstationarity in {Southern Hemisphere} climate variability associated with
  the seasonal breakdown of the stratospheric polar vortex. \textit{J.\
  Climate}, \textbf{30~(18)}, 7125--7139.

\bibitem[{Chen and Plumb(2009)Chen, and Plumb}]{chen2009quantifying}
Chen, G., and R.~A. Plumb, 2009: Quantifying the eddy feedback and the
  persistence of the zonal index in an idealized atmospheric model. \textit{J.\
  Atmos.\ Sci.}, \textbf{66~(12)}, 3707--3720.

\bibitem[{Chen and Zurita-Gotor(2008)Chen, and
  Zurita-Gotor}]{chen2008tropospheric}
Chen, G., and P.~Zurita-Gotor, 2008: The tropospheric jet response to
  prescribed zonal forcing in an idealized atmospheric model. \textit{J.\
  Atmos.\ Sci.}, \textbf{65~(7)}, 2254--2271.

\bibitem[{Eliassen(1951)}]{eliassen1951slow}
Eliassen, A., 1951: Slow thermally or frictionally controlled meridional
  circulation in a circular vortex. \textit{Astrophys. Norveg.}, \textbf{5},
  19--60.

\bibitem[{Feldstein and Lee(1998)Feldstein, and Lee}]{feldstein1998atmospheric}
Feldstein, S., and S.~Lee, 1998: Is the atmospheric zonal index driven by an
  eddy feedback? \textit{J.\ Atmos.\ Sci.}, \textbf{55~(19)}, 3077--3086.

\bibitem[{Feldstein(2000)}]{feldstein2000timescale}
Feldstein, S.~B., 2000: The timescale, power spectra, and climate noise
  properties of teleconnection patterns. \textit{J.\ Climate},
  \textbf{13~(24)}, 4430--4440.

\bibitem[{Fuchs et~al.(2015)Fuchs, Sherwood,, and
  Hernandez}]{fuchs2015exploration}
Fuchs, D., S.~Sherwood, and D.~Hernandez, 2015: An exploration of multivariate
  fluctuation dissipation operators and their response to sea surface
  temperature perturbations. \textit{J.\ Atmos.\ Sci.}, \textbf{72~(1)},
  472--486.

\bibitem[{Gerber et~al.(2008{\natexlab{a}})Gerber, Polvani,, and
  Ancukiewicz}]{gerber2008annular}
Gerber, E.~P., L.~M. Polvani, and D.~Ancukiewicz, 2008{\natexlab{a}}: Annular
  mode time scales in the intergovernmental panel on climate change fourth
  assessment report models. \textit{Geophys.\ Res.\ Lett.}, \textbf{35~(22)}.

\bibitem[{Gerber and Vallis(2007)Gerber, and Vallis}]{gerber2007eddy}
Gerber, E.~P., and G.~K. Vallis, 2007: Eddy--zonal flow interactions and the
  persistence of the zonal index. \textit{J.\ Atmos.\ Sci.}, \textbf{64~(9)},
  3296--3311.

\bibitem[{Gerber et~al.(2008{\natexlab{b}})Gerber, Voronin,, and
  Polvani}]{gerber2008testing}
Gerber, E.~P., S.~Voronin, and L.~M. Polvani, 2008{\natexlab{b}}: Testing the
  annular mode autocorrelation time scale in simple atmospheric general
  circulation models. \textit{Mon.\ Wea.\ Rev.}, \textbf{136~(4)}, 1523--1536.

\bibitem[{Hartmann and Lo(1998)Hartmann, and Lo}]{hartmann1998wave}
Hartmann, D.~L., and F.~Lo, 1998: Wave-driven zonal flow vacillation in the
  {Southern H}emisphere. \textit{J.\ Atmos.\ Sci.}, \textbf{55~(8)},
  1303--1315.

\bibitem[{Hassanzadeh and Kuang(2015)Hassanzadeh, and Kuang}]{pedram15}
Hassanzadeh, P., and Z.~Kuang, 2015: Blocking variability: {Arctic
  Amplification versus Arctic Oscillation}. \textit{Geophys.\ Res.\ Lett.},
  \textbf{42}.

\bibitem[{Hassanzadeh and Kuang(2016{\natexlab{a}})Hassanzadeh, and
  Kuang}]{hassanzadeh2016linear}
Hassanzadeh, P., and Z.~Kuang, 2016{\natexlab{a}}: The linear response function
  of an idealized atmosphere. {Part I: Construction using Green's} functions
  and applications. \textit{J.\ Atmos.\ Sci.}, \textbf{73~(9)}, 3423--3439.

\bibitem[{Hassanzadeh and Kuang(2016{\natexlab{b}})Hassanzadeh, and
  Kuang}]{hassanzadeh2016linear2}
Hassanzadeh, P., and Z.~Kuang, 2016{\natexlab{b}}: The linear response function
  of an idealized atmosphere. {Part II}: Implications for the practical use of
  the fluctuation--dissipation theorem and the role of operator's nonnormality.
  \textit{J.\ Atmos.\ Sci.}, \textbf{73~(9)}, 3441--3452.

\bibitem[{Hassanzadeh et~al.(2014)Hassanzadeh, Kuang,, and Farrell}]{pedram14}
Hassanzadeh, P., Z.~Kuang, and B.~F. Farrell, 2014: Responses of midlatitude
  blocks and wave amplitude to changes in the meridional temperature gradient
  in an idealized dry {GCM}. \textit{Geophys.\ Res.\ Lett.}, \textbf{41~(14)}.

\bibitem[{Held(2005)}]{held2005gap}
Held, I.~M., 2005: The gap between simulation and understanding in climate
  modeling. \textit{Bull.\ Amer.\ Meteor.\ Soc.}, \textbf{86~(11)}, 1609--1614.

\bibitem[{Held and Suarez(1994)Held, and Suarez}]{held1994proposal}
Held, I.~M., and M.~J. Suarez, 1994: A proposal for the intercomparison of the
  dynamical cores of atmospheric general circulation models. \textit{Bull.\
  Amer.\ Meteor.\ Soc.}, \textbf{75~(10)}, 1825--1830.

\bibitem[{James(1987)}]{james1987suppression}
James, I.~N., 1987: Suppression of baroclinic instability in horizontally
  sheared flows. \textit{J.\ Atmos.\ Sci.}, \textbf{44~(24)}, 3710--3720.

\bibitem[{James and Gray(1986)James, and Gray}]{james1986concerning}
James, I.~N., and L.~J. Gray, 1986: Concerning the effect of surface drag on
  the circulation of a baroclinic planetary atmosphere. \textit{Quart.\ J.\
  Roy.\ Meteor.\ Soc.}, \textbf{112~(474)}, 1231--1250.

\bibitem[{Jeevanjee et~al.(2017)Jeevanjee, Hassanzadeh, Hill,, and
  Sheshadri}]{jeevanjee2017perspective}
Jeevanjee, N., P.~Hassanzadeh, S.~Hill, and A.~Sheshadri, 2017: A perspective
  on climate model hierarchies. \textit{J. Adv. Model. Earth Syst.},
  \textbf{9}, 1760--1771.

\bibitem[{Khodkar and Hassanzadeh(2018)Khodkar, and
  Hassanzadeh}]{khodkar2018data}
Khodkar, M.~A., and P.~Hassanzadeh, 2018: Data-driven reduced modelling of
  turbulent {Rayleigh-B\'enard} convection using {DMD}-enhanced
  {Fluctuation-Dissipation Theorem}. \textit{J. Fluid Mech.}, \textbf{852}.

\bibitem[{Khodkar et~al.(2018)Khodkar, Hassanzadeh, Nabi,, and
  Grover}]{khodkar2018reduced}
Khodkar, M.~A., P.~Hassanzadeh, S.~Nabi, and P.~Grover, 2018: Reduced-order
  modeling of fully turbulent buoyancy-driven flows using the {G}reen's
  function method. \textit{arXiv:1805.01596 (Phys. Rev. Fluids, in revision)}.

\bibitem[{Kidson(1988)}]{kidson1988indices}
Kidson, J.~W., 1988: Indices of the {Southern H}emisphere zonal wind.
  \textit{J.\ Climate}, \textbf{1~(2)}, 183--194.

\bibitem[{Kuang(2010)}]{kuang2010linear}
Kuang, Z., 2010: Linear response functions of a cumulus ensemble to temperature
  and moisture perturbations and implications for the dynamics of convectively
  coupled waves. \textit{J.\ Atmos.\ Sci.}, \textbf{67~(4)}, 941--962.

\bibitem[{Kuang(2012)}]{kuang2012weakly}
Kuang, Z., 2012: Weakly forced mock {Walker} cells. \textit{J.\ Atmos.\ Sci.},
  \textbf{69~(9)}, 2759--2786.

\bibitem[{Kuroda and Mukougawa(2011)Kuroda, and Mukougawa}]{kuroda2011role}
Kuroda, Y., and H.~Mukougawa, 2011: Role of medium-scale waves on the {Southern
  Annular Mode}. \textit{J.\ Geophys.\ Res.}, \textbf{116~(D22)}.

\bibitem[{Limpasuvan and Hartmann(1999)Limpasuvan, and
  Hartmann}]{limpasuvan1999eddies}
Limpasuvan, V., and D.~L. Hartmann, 1999: Eddies and the annular modes of
  climate variability. \textit{Geophys.\ Res.\ Lett.}, \textbf{26~(20)},
  3133--3136.

\bibitem[{Limpasuvan and Hartmann(2000)Limpasuvan, and
  Hartmann}]{limpasuvan2000wave}
Limpasuvan, V., and D.~L. Hartmann, 2000: Wave-maintained annular modes of
  climate variability. \textit{J.\ Climate}, \textbf{13~(24)}, 4414--4429.

\bibitem[{Liu et~al.(2018)Liu, Lu, Garuba, Leung, Luo,, and
  Wan}]{liu2018sensitivity}
Liu, F., J.~Lu, O.~Garuba, L.~R. Leung, Y.~Luo, and X.~Wan, 2018: Sensitivity
  of surface temperature to oceanic forcing via q-flux {G}reen's function
  experiments. part i: Linear response function. \textit{J.\ Climate},
  \textbf{31~(9)}, 3625--3641.

\bibitem[{Lorenz(2014)}]{lorenz2014understanding}
Lorenz, D.~J., 2014: Understanding midlatitude jet variability and change using
  {R}ossby wave chromatography: Wave--mean flow interaction. \textit{J.\
  Atmos.\ Sci.}, \textbf{71~(10)}, 3684--3705.

\bibitem[{Lorenz and Hartmann(2001)Lorenz, and Hartmann}]{lorenz2001eddy}
Lorenz, D.~J., and D.~L. Hartmann, 2001: Eddy-zonal flow feedback in the
  {Southern Hemisphere}. \textit{J.\ Atmos.\ Sci.}, \textbf{58~(21)},
  3312--3327.

\bibitem[{Lorenz and Hartmann(2003)Lorenz, and Hartmann}]{lorenz2003eddy}
Lorenz, D.~J., and D.~L. Hartmann, 2003: Eddy--zonal flow feedback in the
  {Northern H}emisphere winter. \textit{J.\ Climate}, \textbf{16~(8)},
  1212--1227.

\bibitem[{Lutsko et~al.(2015)Lutsko, Held,, and
  Zurita-Gotor}]{lutsko2015applying}
Lutsko, N.~J., I.~M. Held, and P.~Zurita-Gotor, 2015: Applying the
  fluctuation-dissipation theorem to a two-layer model of quasi-geostrophic
  turbulence. \textit{J.\ Atmos.\ Sci.}, \textbf{72}, 3161--3177.

\bibitem[{Ma et~al.(2017)Ma, Hassanzadeh,, and Kuang}]{ma2017quantifying}
Ma, D., P.~Hassanzadeh, and Z.~Kuang, 2017: Quantifying the eddy--jet feedback
  strength of the annular mode in an idealized {GCM} and reanalysis data.
  \textit{J.\ Atmos.\ Sci.}, \textbf{74~(2)}, 393--407.

\bibitem[{Nakamura(1993)}]{nakamura1993momentum}
Nakamura, N., 1993: Momentum flux, flow symmetry, and the nonlinear barotropic
  governor. \textit{J.\ Atmos.\ Sci.}, \textbf{50~(14)}, 2159--2179.

\bibitem[{Nie et~al.(2014)Nie, Zhang, Chen, Yang,, and
  Burrows}]{nie2014quantifying}
Nie, Y., Y.~Zhang, G.~Chen, X.-Q. Yang, and D.~A. Burrows, 2014: Quantifying
  barotropic and baroclinic eddy feedbacks in the persistence of the {Southern
  Annular Mode}. \textit{Geophys.\ Res.\ Lett.}, \textbf{41~(23)}, 8636--8644.

\bibitem[{Nigam(1990)}]{nigam1990structure}
Nigam, S., 1990: On the structure of variability of the observed tropospheric
  and stratospheric zonal-mean zonal wind. \textit{J.\ Atmos.\ Sci.},
  \textbf{47~(14)}, 1799--1813.

\bibitem[{Ring and Plumb(2008)Ring, and Plumb}]{ring08}
Ring, M.~J., and R.~A. Plumb, 2008: The response of a simplified {GCM} to
  axisymmetric forcings: Applicability of the fluctuation-dissipation theorem.
  \textit{J.\ Atmos.\ Sci.}, \textbf{65~(12)}, 3880--3898.

\bibitem[{Rivi{\`e}re et~al.(2016)Rivi{\`e}re, Robert,, and
  Codron}]{riviere2016short}
Rivi{\`e}re, G., L.~Robert, and F.~Codron, 2016: A short-term negative eddy
  feedback on midlatitude jet variability due to planetary wave reflection.
  \textit{J.\ Atmos.\ Sci.}, \textbf{73~(11)}, 4311--4328.

\bibitem[{Robert et~al.(2017)Robert, Rivi{\`e}re,, and
  Codron}]{robert2017positive}
Robert, L., G.~Rivi{\`e}re, and F.~Codron, 2017: Positive and negative eddy
  feedbacks acting on midlatitude jet variability in a three-level
  quasigeostrophic model. \textit{J.\ Atmos.\ Sci.}, \textbf{74~(5)},
  1635--1649.

\bibitem[{Robinson(1991)}]{robinson1991dynamics}
Robinson, W.~A., 1991: The dynamics of low-frequency variability in a simple
  model of the global atmosphere. \textit{J.\ Atmos.\ Sci.}, \textbf{48~(3)},
  429--441.

\bibitem[{Robinson(1996)}]{robinson1996does}
Robinson, W.~A., 1996: Does eddy feedback sustain variability in the zonal
  index? \textit{J.\ Atmos.\ Sci.}, \textbf{53~(23)}, 3556--3569.

\bibitem[{Robinson(2000)}]{robinson2000baroclinic}
Robinson, W.~A., 2000: A baroclinic mechanism for the eddy feedback on the
  zonal index. \textit{J.\ Atmos.\ Sci.}, \textbf{57~(3)}, 415--422.

\bibitem[{Ronalds et~al.(2018)Ronalds, Barnes,, and
  Hassanzadeh}]{ronalds2018barotropic}
Ronalds, B., E.~Barnes, and P.~Hassanzadeh, 2018: A barotropic mechanism for
  the response of jet stream variability to {Arctic Amplification} and sea ice
  loss. \textit{J.\ Climate}, \textbf{31}, 7069--7085.

\bibitem[{Sato et~al.(2000)Sato, Yamada,, and Hirota}]{sato2000global}
Sato, K., K.~Yamada, and I.~Hirota, 2000: Global characteristics of
  medium-scale tropopausal waves observed in {ECMWF} operational data.
  \textit{Mon.\ Wea.\ Rev.}, \textbf{128~(11)}, 3808--3823.

\bibitem[{Shepherd(2014)}]{shepherd2014}
Shepherd, T.~G., 2014: Atmospheric circulation as a source of uncertainty in
  climate change projections. \textit{Nat. Geosci.}, \textbf{7}, 703--708.

\bibitem[{Sheshadri and Plumb(2017)Sheshadri, and
  Plumb}]{sheshadri2017propagating}
Sheshadri, A., and R.~A. Plumb, 2017: Propagating annular modes: Empirical
  orthogonal functions, principal oscillation patterns, and time scales.
  \textit{J.\ Atmos.\ Sci.}, \textbf{74~(5)}, 1345--1361.

\bibitem[{Sheshadri et~al.(2018)Sheshadri, Plumb, Lindgren,, and
  Domeisen}]{sheshadri2018vertical}
Sheshadri, A., R.~A. Plumb, E.~A. Lindgren, and D.~I.~V. Domeisen, 2018: The
  vertical structure of annular modes. \textit{J.\ Atmos.\ Sci.}

\bibitem[{Simpson et~al.(2013)Simpson, Shepherd, Hitchcock,, and
  Scinocca}]{simpson2013southernII}
Simpson, I.~R., T.~G. Shepherd, P.~Hitchcock, and J.~F. Scinocca, 2013:
  Southern annular mode dynamics in observations and models. part {II}: Eddy
  feedbacks. \textit{J.\ Climate}, \textbf{26~(14)}, 5220--5241.

\bibitem[{Thompson and Barnes(2014)Thompson, and Barnes}]{thompson2014periodic}
Thompson, D. W.~J., and E.~A. Barnes, 2014: Periodic variability in the
  large-scale {Southern Hemisphere} atmospheric circulation. \textit{Science},
  \textbf{343~(6171)}, 641--645.

\bibitem[{Thompson and Li(2015)Thompson, and Li}]{thompson2015baroclinic}
Thompson, D. W.~J., and Y.~Li, 2015: Baroclinic and barotropic annular
  variability in the {Northern Hemisphere}. \textit{J.\ Atmos.\ Sci.},
  \textbf{72~(3)}, 1117--1136.

\bibitem[{Thompson and Wallace(1998)Thompson, and Wallace}]{thompson1998arctic}
Thompson, D. W.~J., and J.~M. Wallace, 1998: The {Arctic Oscillation} signature
  in the wintertime geopotential height and temperature fields.
  \textit{Geophys.\ Res.\ Lett.}, \textbf{25~(9)}, 1297--1300.

\bibitem[{Thompson and Woodworth(2014)Thompson, and
  Woodworth}]{thompson2014barotropic}
Thompson, D. W.~J., and J.~D. Woodworth, 2014: Barotropic and baroclinic
  annular variability in the {Southern Hemisphere}. \textit{J.\ Atmos.\ Sci.},
  \textbf{71~(4)}, 1480--1493.

\bibitem[{Vallis et~al.(2004)Vallis, Gerber, Kushner,, and
  Cash}]{vallis2004mechanism}
Vallis, G.~K., E.~P. Gerber, P.~J. Kushner, and B.~A. Cash, 2004: A mechanism
  and simple dynamical model of the {North Atlantic Oscillation} and annular
  modes. \textit{J.\ Atmos.\ Sci.}, \textbf{61~(3)}, 264--280.

\bibitem[{Yu and Hartmann(1993)Yu, and Hartmann}]{yu1993zonal}
Yu, J.-Y., and D.~L. Hartmann, 1993: Zonal flow vacillation and eddy forcing in
  a simple gcm of the atmosphere. \textit{J.\ Atmos.\ Sci.}, \textbf{50~(19)},
  3244--3259.

\bibitem[{Zurita-Gotor et~al.(2014)Zurita-Gotor, Blanco-Fuentes,, and
  Gerber}]{zurita2014impact}
Zurita-Gotor, P., J.~Blanco-Fuentes, and E.~P. Gerber, 2014: The impact of
  baroclinic eddy feedback on the persistence of jet variability in the
  two-layer model. \textit{J.\ Atmos.\ Sci.}, \textbf{71~(1)}, 410--429.

\end{thebibliography}



\end{document}